\documentclass[twocolumn,twoside]{IEEEtran}

\ifCLASSINFOpdf

\else

\fi

\ifCLASSOPTIONcompsoc
    \usepackage[caption=false, font=normalsize, labelfont=sf, textfont=sf]{subfig}
\else
    \usepackage[caption=false, font=normalsize]{subfig}
\fi
\usepackage{lipsum}%
\usepackage[dvipsnames]{xcolor}
\usepackage{algorithm,algorithmic}
\usepackage{balance}
\usepackage{multicol}   
\usepackage{cite}
\usepackage{gensymb}
\usepackage{multirow}
\usepackage{graphics}
\usepackage{epsfig}
\usepackage{graphicx}
\usepackage{epstopdf}
\usepackage{textcomp}
\usepackage{amsmath}
\usepackage{mathtools}
\usepackage{filecontents}
\usepackage{lipsum,color}
\usepackage{amssymb}
\usepackage{float}
\usepackage{colortbl} 
\usepackage{times} 
\usepackage{amsthm}  

\usepackage{amsfonts}

\theoremstyle{break}

\begin{document}
\title{Enabling Long mmWave Aerial Backhaul Links via Fixed-Wing UAVs: 
	Performance and Design}

\author{ Mohammad~Taghi~Dabiri,~Mazen~Omar~Hasna,~{\it Senior Member,~IEEE},
	~Nizar~Zorba,~{\it Senior Member,~IEEE},
	~Tamer~Khattab,~{\it Senior Member,~IEEE}
	%
	\thanks{The authors are with the Department of Electrical Engineering, Qatar University, Doha, Qatar.  (E-mail: m.dabiri@qu.edu.qa; hasna@qu.edu.qa; nizarz@qu.edu.qa; tkhattab@ieee.org).}
}

\maketitle
\vspace{-1cm}
\begin{abstract}
We propose a fixed wing unmanned aerial vehicles (UAV)-based millimeter wave (mmWave) backhaul links, that is offered as a cost effective and easy to deploy solution,  to connect a disaster or remote area to the nearest core network. First, we fully characterize the single relay fixed-wing UAV-based communication system by taking into account the effects of realistic physical parameters, such as the UAV's circular path, critical points of the flight path,
heights and positions of obstacles, flight altitude, tracking error, the severity of UAV's vibrations, the real 3D antenna pattern, mmWave atmospheric channel loss, temperature and air pressure. Second, we derive the distribution of the signal-to-noise ratio (SNR) metric, which is based on the sum of a series of  Dirac delta functions. Using the SNR distribution, we derive closed-form expressions for the outage probability and the ergodic capacity of the considered system as a function of all system parameters. To provide an acceptable quality of service for longer link lengths, we extend the analytical expressions to a multi-relay system. 
The accuracy of the closed-form expressions are verified by Monte-Carlo simulations.
Finally, by providing sufficient simulation results, we investigate the effects of key channel parameters such as antenna pattern gain and flight path on the performance of the considered system; and we carefully analyze the relationships between these parameters in order to maximize the average channel capacity.
\end{abstract}
\IEEEpeerreviewmaketitle

\section{Introduction}
\subsection{Background}



\IEEEPARstart{C}{limate} change has been the main case for severe storms, flooding and hurricanes in the recent years. 
Over the past three decades, Europe has seen a sixty percent increase in extreme weather events \cite{web2019}.
Over the past three years, the average number of billion-dollar disasters in the US was more than double the long-term average \cite{web2020}.
%
Parts of the world that have never experienced severe weather should be ready and plan for it now, while those who are more used to these extreme natural events should be prepared for more \cite{web2019}.
One of the essential needs during and after a disaster event is providing a reliable connection link quickly to facilitate rescue operations, as well as to provide internet connectivity to the people escaping from the affected area \cite{ullah2020cognition}.
Therefore, immediate and cost efficient high throughput solutions must be considered after natural disasters even better and more ubiquitous for 5G evolution and beyond.

Natural disasters comprising earthquakes, hurricanes, tornadoes, floods, and other geologic processes can potentially cut or entirely destroy fiber infrastructure to the disaster area.
Any disruption to the fragile fiber causes data disconnections that take days to find and repair.
On the other hands, providing an alternative terrestrial wireless backhaul connectivity encounters serious challenges, including creating a line of sight (LoS) between the disaster area to the nearest core network, especially in forest and mountainous areas \cite{galkin2018backhaul}.
Due to their unique capabilities such as flexibility, maneuverability,  and adaptive altitude adjustment, unmanned aerial vehicles (UAVs) acting as networked flying platforms (NFPs) can be considered as a promising solution to provide a temporary wireless backhaul connectivity while improving reliability of backhaul operations \cite{alzenad2018fso,khawaja2019survey,saeed2021point}.
More recently, millimeter wave (mmWave) backhauling has been proposed as a promising approach for aerial communications because of three reasons. First, unlike terrestrial mmWave communication links that suffer from blockage, the flying nature of UAVs offers a higher probability of LoS between communication nodes. Second, the large available bandwidth at mmWave frequencies can provide high data rate point-to-point aerial communication links, as needed for the backhaul communications. Third, to mitigate the negative effects of the high path-loss at the mmWave bands, the small wavelength enables the realization of a compact form of highly directive antenna arrays, which are suitable for small UAVs with limited payload.

\subsection{Literature Review and Motivation}
More recently, UAV-based mmWave backhaul links have been studied in \cite{ 9712177,9685132,    9714216,9685565,
	gapeyenko2018flexible, 
	tafintsev2020aerial, 9411710, cicek2020backhaul, feng2018spectrum,dabiri2019analytical,dabiri20203d,dabiri2022study,selim2018post, yu2019uav}.
For instance, a 3D two-hop scheme is proposed in \cite{9712177,9685132} wherein a user is connected to a base station (BS) by using a UAV-based backhaul link. 
In particular, the authors studied the performance of the considered network in both amplify-andforward (AF) and decode-and-forward (DF) relaying protocols by considering realistic antenna radiation patterns for both BSs and UAVs based on practical models developed by 3GPP.
A novel wireless backhaul link is suggested in \cite{9714216,9685565} by installing reconfigurable intelligent surface (RIS) on high altitude UAVs to handle a sudden increase in traffic in an urban area.
In \cite{gapeyenko2018flexible,tafintsev2020aerial}, the authors investigated the use of aerial relay node to provide a flexible and reconfigurable backhaul architecture by considering the effects of multipath propagation and dynamic link blockage in mmWave frequency bands.
In \cite{9411710}, the achievable rate of the UAV-based mmWave wireless backhaul link is investigated and then, the authors analyzed the minimum cache hit probability to achieve a certain backhaul rate requirement. 
The UAV-BS location and bandwidth allocation problems is studied in \cite{cicek2020backhaul} to maximize the throughput without exceeding the backhaul and access capacities.
A novel spectrum management architecture for UAV-assisted mmWave networks is suggested in \cite{feng2018spectrum} to overcome the problem related to the spatio-temporal distribution of the wireless network traffic. In particular, with numerical results, the authors studied the performances of the proposed spectrum management for mmWave based backhaul in five different scenarios.
The performance of UAV-based mmWave backhaul link is investigated in \cite{dabiri2019analytical,dabiri20203d} when UAVs are equipped with linear and square array antennas.
More recently, a fast algorithm for 3D optimal placement of rotary-wing UAVs is proposed in \cite{dabiri2022study} to provide a long  mmWave backhaul link.
In \cite{yu2019uav}, the authors studied a UAV-aided low latency mobile edge computing network with mmWave backhauling. 
However, the results of these studies are limited for rotary-wing UAVs.

Rotary wing UAVs are used in cases where more maneuverability is required, for example, to provide internet service in crowded urban areas.
To keep the rotary wing UAV stable in the air, its motors are required to individually speed up or slow down its propellers, which can be time consuming, mainly due to UAV inertia. Moreover, scaling the rotary-wing UAV up to a larger size faces major challenges because more energy is needed to change the speed of larger propellers. 
They also face restrictions on payload, altitude, and shorter flight times.
%
Being able to fly for longer times, at higher altitudes, and with heavier payloads than rotary-wing UAVs are the greatest advantages of fixed wing UAVs.
All these characteristics make them suitable for remote or disaster area applications. 
%
Based on the results of \cite{dabiri2019analytical,dabiri20203d}, to design an aerial mmWave backhaul link based on a rotary wing UAV, it is needed to find an optimal point in 3D space relative to the ground transmitter and receiver.
%
However, fixed wing UAVs cannot hover or make sharp turns, and thus, the results of the aforementioned works are not directly applicable for fixed wing UAVs.


\subsection{Contributions and Paper Structure}
\begin{figure}
	\centering
	\subfloat[] {\includegraphics[width=3.3 in]{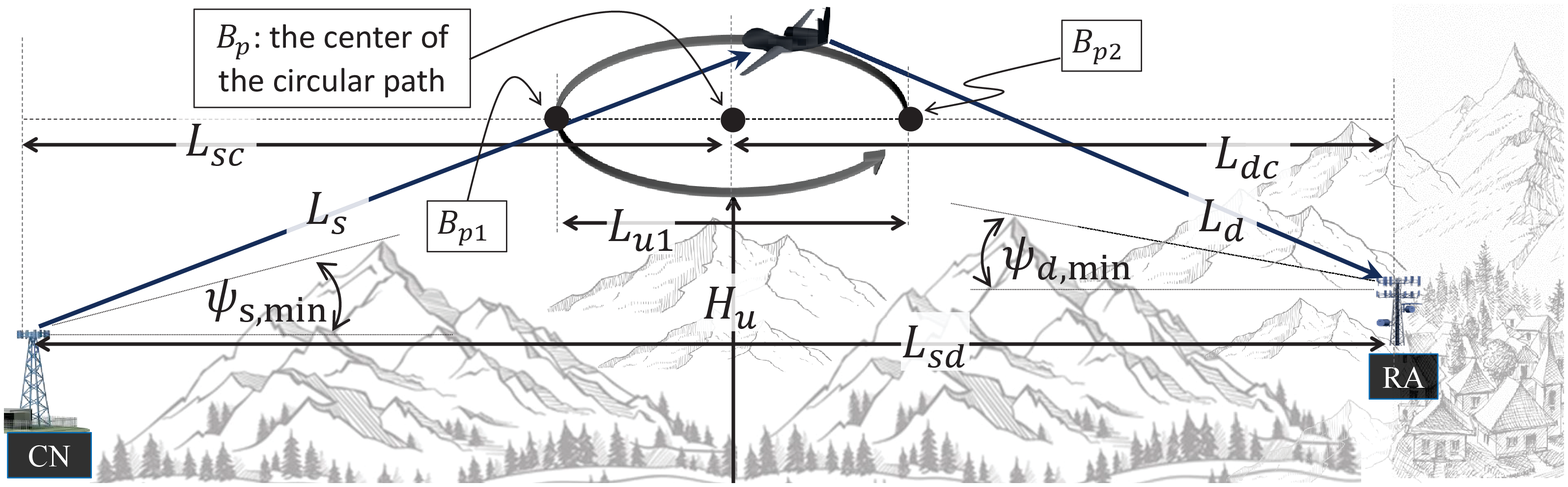}
		\label{n1}
	}
	\hfill
	\subfloat[] {\includegraphics[width=3.3 in]{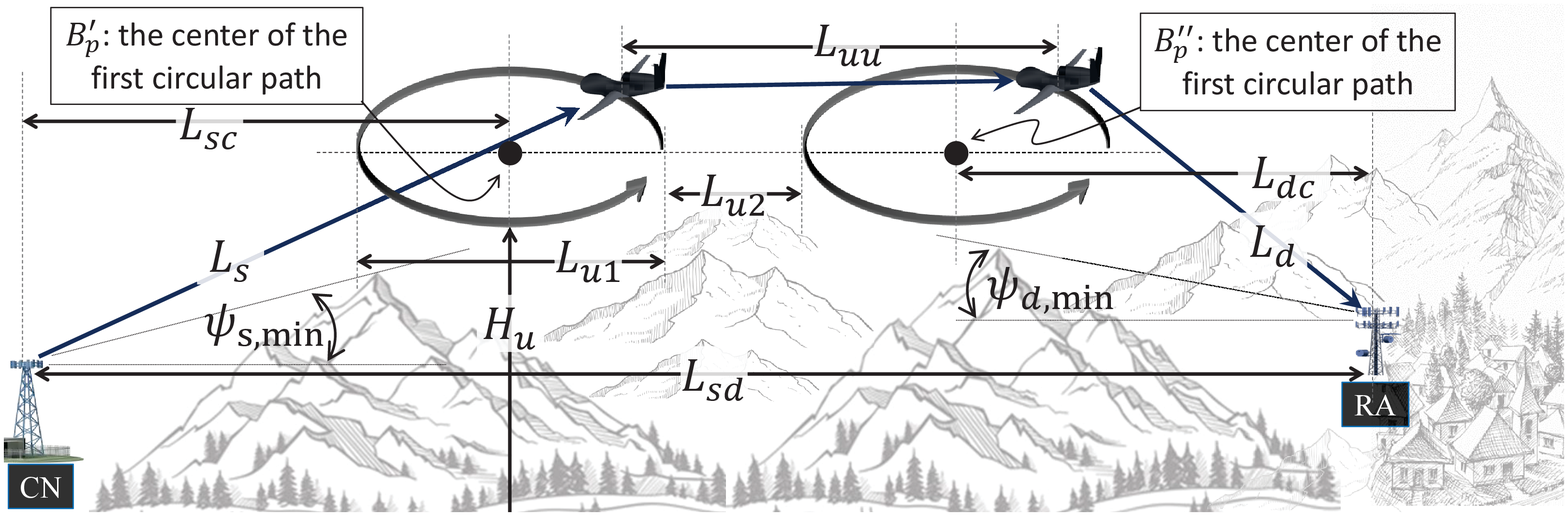}
		\label{n2}
	}
	\caption{A fixed wing UAV acting as an NFP node in order to relay data from the nearest core network to the disaster or remote area: (a) for single relay topology and (b) two relay topology.}
	\label{n3}
\end{figure}

In this study, we consider a mmWave  backhaul link based on fixed wing UAVs, as shown in Fig. \ref{n3}, that is offered as a cost effective and easy to deploy solution to connect a disaster or remote area to the nearest core network in a short time.
Performance analysis and optimal parameters for system design of the considered fixed-wing UAV-based communication system are the main contributions of this work by taking into account the realistic parameters. 
Our detailed contributions are summarized as follows:

\begin{itemize}
	\item We fully characterize the performance of single relay fixed-wing UAV-based communication system by taking into account the effects of realistic physical parameters, such as the UAV's circular path, critical points of the flight path,
	heights and position of obstacles, flight altitude, tracking error, severity of UAV's vibrations, real 3D antenna pattern, mmWave atmospheric channel loss, temperature and air pressure.
	\item 
	We derive the distribution of the signal-to-noise ratio (SNR), showing that the distribution of end-to-end SNR corresponds to the sum of a series of  Dirac delta functions.
	\item We derive closed-form expressions for the outage probability and channel capacity of the considered system, as a function of considered practical system parameters.
	The accuracy of closed-form expressions is verified with the results obtained from Monte-Carlo simulations. 
	The main feature of the provided analytical expressions is that they are function of all key channel parameters, showing the impact of each parameter on the system performance.
	%
	%
	\item Through extensive simulation results, we show the effects of key channel parameters such as antenna pattern gain and optimal flight path on the performance of the considered system, and we carefully study the relationships between those parameters in order to maximize average channel capacity.
	\item Within some scenarios, the use of single relay UAV will not be able to guarantee the requested quality of service (QoS) for longer links. To provide an acceptable QoS for longer link length, we extend the analytical expressions to a multi-relay system. Then, using the obtained analytical expressions, we study the optimal parameter design for a multi-relay system.
\end{itemize}

The rest of this paper is organized as follows. We introduce the channel model of a fixed wing UAV-based mmWave backhaul link in Section II.
Analytical derivations along with the performance analysis of the considered system, in terms of the channel capacity and the outage probability are provided in Section III.
Using the numerical and simulation results, we study the optimal parameters design of the considered system in section IV.
Finally, conclusions and future road map are drawn in Section V.

%
\begin{figure}
	\begin{center}
		\includegraphics[width=2.8 in]{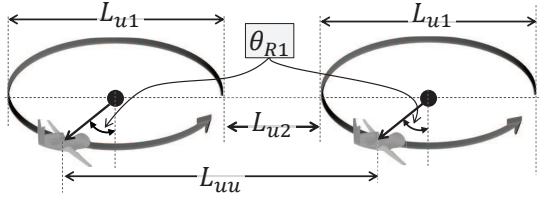}
		\caption{An illustration of the coordinated flight of the UAVs in a circular path so that the distance between the UAVs is constant and equal to $L_{uu}$ during the whole circular flight path.}
		\label{rn1}
	\end{center}
\end{figure}
%

\begin{table}
	\caption{List of main notations.} 
	\centering 
	\begin{tabular}{l l} 
		\hline\hline \\[-1.2ex]
		{\bf Parameter} & {\bf Description}  \\ [.5ex] 
		\hline\hline \\[-1.2ex]
		\multicolumn{2}{l}{ \textbf{{Single UAV}} } \\[-1ex]
		\multicolumn{2}{l}{ ------------------------------ } \\ 
		$v\in\{t,r\}$ & This subscript is used to specify Tx and Rx antennas \\
		$q\in\{s,d\}$ & This subscript is used to specify $A_s$ and $A_d$ \\
		$w\in\{x,y\}$ & This subscript is used to specify $x$ and $y$ axes \\
		$A_{us}$ & The NFP antenna directed towards the $A_s$ \\
		$A_{ud}$ & The NFP antenna directed towards the $A_d$  \\
		$A_s$ & Antenna of CN\\
		$A_d$ & Antenna of RA\\
		$P_{t,s}$ & Transmitted power of $A_s$\\
		$P_{t,d}$ & Transmitted power of $A_{ut}$\\
		$N_{uqw}$ & Number of antenna elements of $A_{uq}$ in $w_q$ axis \\
		$N_{qw}$ & Number of antenna elements of $A_q$ in $w_q$ axis \\
		$\theta_{qw}$    & Instantaneous misalignment of $A_q$ in $w_q-z_q$ plane \\
		$\theta_{uqw}$    & Instantaneous misalignment of $A_{uv}$ in $w_q-z_q$ plane \\
		$\mu_{qw}$    &  Mean  of RV $\theta_{qw}$ \\
		$\sigma_{qw}^2$    &  Variance of RV $\theta_{qw}$ \\
		$\mu_{uqw}$    &  Mean of RV $\theta_{uqw}$ \\
		$\sigma_{uqw}^2$    &  Variance of RV $\theta_{uqw}$ \\
		$\lambda$ and $f_c$ &  Wavelength and carrier frequency, respectively \\
		$B_{p1}$ & The farthest and closest point to $A_d$ and $A_s$ \\
		$B_{p2}$ & The farthest and closest point to $A_s$ and $A_d$ \\
		$B_{p}$ & The center of UAV circular path \\
		%
		%
		%
		%
		$\psi_{q,\text{min}}$ & Minimum elevation angle\\
		$H_u$ & Heights of NFP \\
		$L_q$ & Link length of $A_q$ to NFP \\
		$L_{sd}$ & Horizontal distance between $A_s$ and $A_d$ \\
		$L_{qc}$ & Horizontal distance between $A_q$ and point $B_p$ \\  
		$L_{u1}$ & Diameter of the UAV circular flight path \\ 
		$\theta_{R1}$ & Determines the UAV's position in a circular path\\
		\multicolumn{2}{l}{ ------------------------------ } \\
		%
		\multicolumn{2}{l}{ \textbf{{Multiple UAVs}} } \\[-1ex]
		\multicolumn{2}{l}{ ------------------------------ } \\
		$B_{p'}$  & The center of first UAV circular path \\
		$B_{p''}$ & The center of second UAV circular path \\
		$A_{ut1}$ & The first UAV antenna directed toward the $A_{ur2}$ \\
		$A_{ur2}$ & The second UAV antenna directed toward the $A_{ut1}$  \\
		$N_{tw1}$ & Number of antenna elements of $A_{ut1}$ in $w_u$ axis \\
		$N_{rw2}$ & Number of antenna elements of $A_{ur2}$ in $w_u$ axis \\
		$L_{u2}$  & Distance between UAVs' circular paths shown in Fig. \ref{n2} \\
		$L_{uu}$  & Inter UAVs link length \\    
		$\theta_{uw1}$ &  Instantaneous misalignment of $A_{ut1}$ in $w_u-z_u$ plane \\
		$\theta_{uw2}$ &  Instantaneous misalignment of $A_{ur2}$ in $w_u-z_u$ plane \\
		$\mu_{uw1}$    &  Mean  of RV $\theta_{uw1}$ \\
		$\mu_{uw2}$    &  Mean  of RV $\theta_{uw2}$ \\
		$\sigma_{uw1}^2$    &  Variance of RV $\theta_{uw1}$ \\
		$\sigma_{uw2}^2$    &  Variance of RV $\theta_{uw2}$ \\
		$M$       & Number of UAVs \\
		\hline
		$\mathbb{P}_\text{our,tr}$    & Outage threshold \\
		$\Gamma_\text{tr}$    & SNR threshold \\
		\hline \hline              
	\end{tabular}
	\label{I1} 
\end{table}

\section{The System Model}
We consider a fixed-wing UAV acting as an NFP node in order to relay data from the nearest core network (CN) to the disaster or remote area (RA), where $L_{sd}$ shows the distance between CN and RA.
First, we assume that the backhaul link is relayed to the RA by one fixed-wing UAV.
However, we will show that for longer values of $L_{sd}$, using only one UAV can not provide a desired QoS. Therefore, in the second part of this work, a relay system based on two or multiple fixed-wing UAVs is studied.

\subsection{Single Relay System}
%
The fixed-wing UAV rotates in a circular path with center $B_p$ and diameter $L_{u1}$ as depicted in Fig. \ref{n1}.
Point $B_{p1}$ shown in Fig.\,\ref{n1} is the closest and farthest point to the CN and RA, respectively. On the other hand, $B_{p2}$ is the farthest and closest point to the CN and RA, respectively.
Let $\psi_{s,\text{min}}$ and $\psi_{d,\text{min}}$ denote the minimum elevation angles\footnote{The minimum elevation angle is the minimum angle required to establish a LoS between the ground node and the nearest UAV.}  of CN and RA, respectively, $L_s$ represents the link length from CN to UAV (CU), $L_d$ denotes the link length between UAV to RA (UR), and $H_u$ stands for the UAV height.
%
%

\begin{figure}
	\centering
	\subfloat[] {\includegraphics[width=3.3 in]{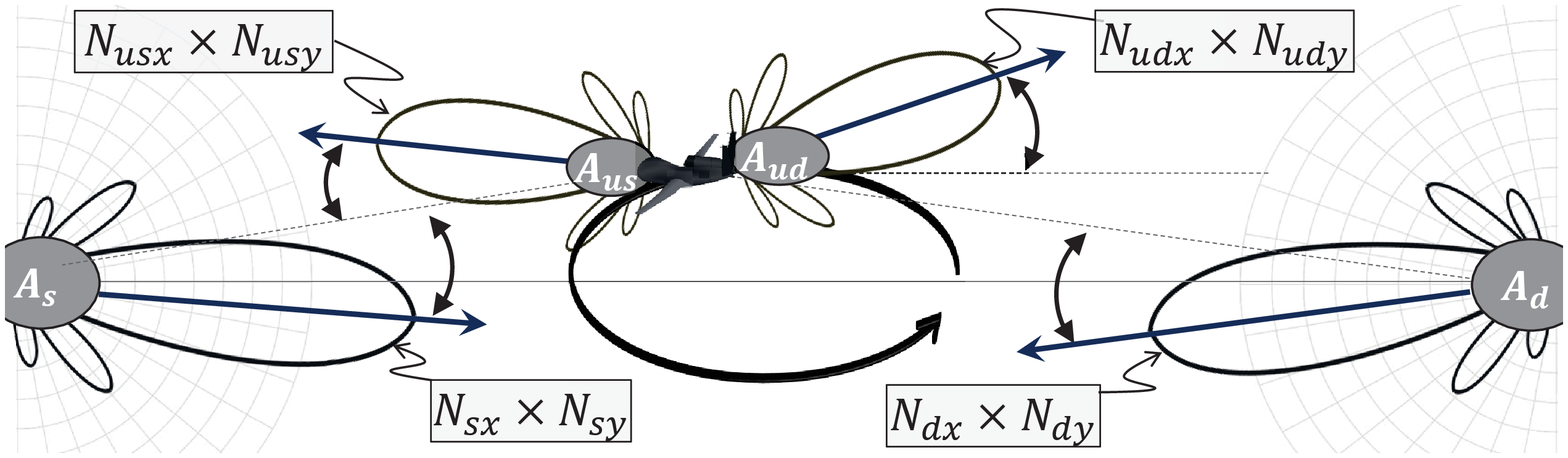}
		\label{m1}
	}
	\hfill
	\subfloat[] {\includegraphics[width=3.3 in]{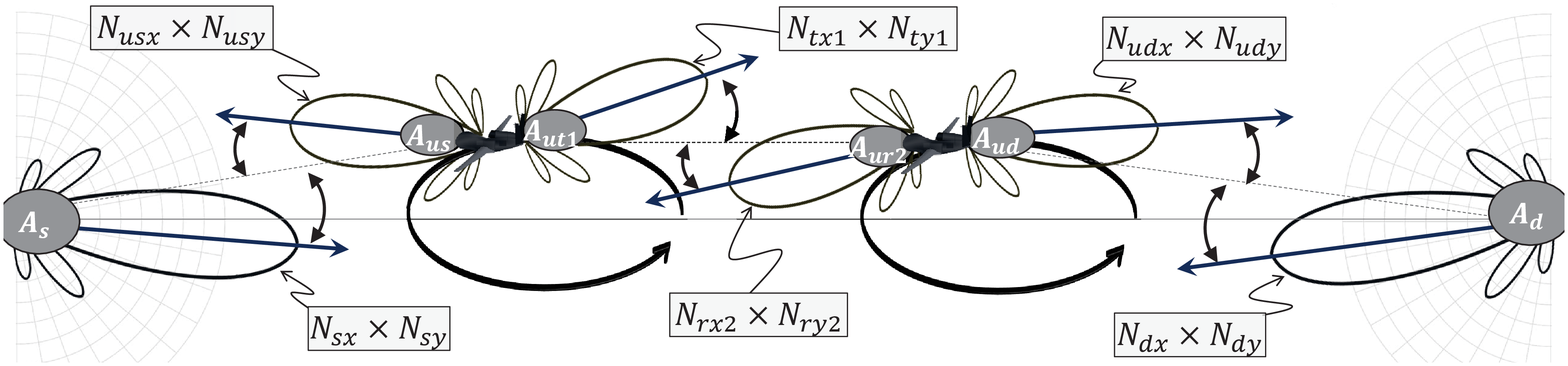}
		\label{m2}
	}
	\caption{A graphical illustration of antenna pattern misalignment: (a) for single relay topology, and (b) for two relay system.}
	\label{m3}
\end{figure}

As shown in Fig. \ref{m1}, the single relay topology consists of four mmWave array antennas.
We consider $z_s$ as the propagation axis of the CU link, while axes $x_s$ and $y_s$ represent the array antenna plane perpendicular to the propagation axis. Similarly, $z_d$ represents the propagation axis of the UD link, while axes $x_d$ and $y_d$ represent the array antenna plane perpendicular to the propagation axis $z_d$.
Let $A_s(N_{sx}\times N_{sy})$ denotes the CN antenna characterized by $N_{sx}\times N_{sy}$ where $N_{sx}$ and $N_{sy}$ are the number of antenna elements in the $x_s-y_s$ plane. Similarly, let $A_d(N_{dx}\times N_{dy})$ denotes the array antenna of RA node, 
$A_{us}(N_{usx}\times N_{usy})$ denotes the array antenna of the NFP directed toward the CN, and $A_{ud}(N_{udx}\times N_{udy})$ denotes the array antenna of NFP directed toward the RA, respectively. 
Antennas $A_s$ and $A_{us}$ as well as antennas $A_d$ and $A_{ud}$ try to adjust the direction of their antennas to each other.
At first it may seem that by increasing the number of antenna elements, which leads to an increase in antenna gain, the system performance improves.
However, in practical situations, increasing the antenna gain makes the system more sensitive to antenna misalignment.
A change in the instantaneous speed and acceleration of the fixed wing UAV,
an error in the mechanical control system of UAV, mechanical noise, position estimation errors, air pressure, and wind speed can cause an alignment error between the antennas  \cite{dabiri2018channel,dabiri2021uav}, as graphically illustrated in Fig. \ref{m1}.
Therefore, the optimal design of the antenna patterns is of great importance in the presence of alignment error.
%
Let 
$\theta_{qw}\sim\mathcal{N}(\mu_{qw},\sigma_{qw}^2)$ be the instantaneous misalignment angle of $A_q$ in $w_q-z_q$ plane, where $q\in\{s,d\}$ and $w\in\{x,y\}$.
Similarly, $\theta_{uqw}\sim\mathcal{N}(\mu_{uqw},\sigma_{uqw}^2)$ is assumed to be the instantaneous misalignment of $A_{uq}$ in $w_q-z_q$ plane.

\subsection{Multiple Relay System}
For the longer values of $L_{sd}$, we must increase the number of UAVs acting as relay to satisfy the required QoS along the entire circular flight path.
The topology of a system with two UAVs is shown in Fig. \ref{n2}.
For notation simplicity, all variables defined for a single-relay system are also valid for a multiple-relay system, except for a few variables that are redefined below.
The mmWave signal first points to the first fixed-wing UAV, which is decoded and forwarded to the second UAV. Similarly, after receiving the signal, the second UAV relay decodes and forwards it to the RA.
Both fixed-wing UAVs rotate in a circular path with diameter $L_{u1}$ as depicted in Figs. \ref{n2} and \ref{rn1}.
The center of circular path for first and second UAVs are $B'_p$ and $B''_p$, respectively.
Both UAVs fly at the same position, at the same speed and in the same direction in a circular path, so that the link length between the two UAVs remains constant along the entire circular flight path, as shown in Fig. \ref{rn1}. The considered link length between the two UAVs  is $L_{uu}=L_{u1}+L_{u2}$, where $L_{u2}$ is the distance between two circular paths. Note that the parameter $L_{u2}$ is a tunable parameter and has a significant effect on the performance of the considered system and thus, finding an optimal value for it is very important.

As shown in Fig. \ref{m2}, the two relay system consists of six mmWave array antennas.
Four of the antennas are similar to the single relay system and the other two antennas are related to UU link.
We consider that $z_u$ represents the propagation axis between the UAVs while axes $x_u$ and $y_u$ represent the array antenna plane perpendicular to the propagation axis. 
As shown in Fig. \ref{m2}, let $A_{ut1}(N_{tx1}\times N_{ty1})$ denote the first UAV antenna directed toward the second UAV which is characterized by $N_{tx1}\times N_{ty1}$. Also, $A_{ur2}(N_{rx1}\times N_{ry1})$ represents the second UAV antenna directed toward the first UAV which is characterized by $N_{rx2}\times N_{ry2}$. 
Antennas $A_{ut1}$ and $A_{ur2}$ try to adjust the direction of their beams to each other.
Let 
$\theta_{uw1}\sim\mathcal{N}(\mu_{uw1},\sigma_{uw1}^2)$ be the instantaneous misalignment angle of $A_{ut1}$ in $w_u-z_u$ plane, $\theta_{uw2}\sim\mathcal{N}(\mu_{uw2},\sigma_{uw2}^2)$ be the instantaneous misalignment angle of $A_{ur2}$ in $w_u-z_u$ plane.

For a multi relay system with $M$ UAVs, we have $M-1$ inter-UAV links. Depending on the type of DF relays used, as well as the symmetry of the inter-UAV links, the optimal parameter values of all links must be the same.
Therefore, the design of a multi-relay system is similar to a two-relay system.

\subsection{Channel Propagation Loss}
In normal atmospheric conditions, water vapor (H$_2$O) and oxygen (O$_2$) molecules are strongly absorptive of radio signals, especially at mmWave frequencies and higher.
%
%
%
%
%
%
The resulting attenuation is in excess of the reduction in radiated signal power due to free-space loss.
Channel loss (in dB) is usually expressed as
\begin{align}
	\label{f1}
	h_{L,\text{dB}}^{\text{tot}}(f_c) = 20\log\left(\frac{4 \pi L }{\lambda}\right)  + h_{L,\text{dB}}^{o,w}(f_c),
\end{align}
where $L$ is the link length (in m), $\lambda$ is the wavelength (in m), $f_c$ is mmWave frequency (in GHz),
$h_{L,\text{dB}}^{o,w}(f_c) = \frac{h_{L,\text{dB/km}}^{o,w}(f_c) L}{1000}$ is the total attenuation due to oxygen and water (in dB), $h_{L,\text{dB/km}}^{o,w}(f_c)=h_{L,\text{dB/km}}^{o}(f_c)+h_{L,\text{dB/km}}^{w}(f_c)$ is the attenuation per km due to oxygen and water (in dB/km).
At 20°C surface temperature and at sea level, approximate expressions for the attenuation constants of oxygen and water vapor (in dB/km)  as defined by the International Telecommunications Union (ITU) are \cite{ITU_1}:
\begin{align}
	\label{po1}
	&h_{L,\text{dB/km}}^{o,0}(f_c) = 0.001\times f_c^2\\
	&\times \left\{
	\begin{array}{rl}
		\frac{6.09}{f_c^2+0.227} + \frac{4.81}{(f_c-57)^2+1.5} ~~~~~~~~~~~& ~~~  f_c<57  \\  
		h_{L,\text{dB/km}}^{o,0}(f_c=57) + 1.5(f_c-57)& ~~~  57<f_c<63 \\
		\frac{4.13}{(f_c-63)^2+1.1} + \frac{0.19}{(f_c-118.7)^2+2} ~~~~~~& ~~~  63<f_c<350
	\end{array} \right. \nonumber
\end{align}
and
\begin{align}
	\label{po2}
	&h_{L,\text{dB/km}}^{w,0}(f_c) = 0.0001\times f_c^2 \rho _0 \left(0.05  + \frac{3.6}{(f_c-22.2)^2+8.5}  \right. \nonumber \\
	& \left. + \frac{10.6}{(f_c-183.3)^2+9}    + \frac{8.9}{(f_c-325.4)^2+26.3} \right), ~~f_c<350,
\end{align}
where $\rho_0=7.5 ~\text{g/m}^3$ is the water vapor density at sea level, and $h_{L,\text{dB/km}}^{o,0}(f_c=57)$ is the value of the first expression at $f_c=57$ GHz.
In general, the attenuation constants of oxygen and water vapor are functions of altitude, since they depend on factors such as temperature and pressure. These quantities are often assumed to vary exponentially with height $H$, as $\rho(H) = \rho_0 \exp\left(-H/H_\text{scale}\right)$ where $H_\text{scale}$ is known as the scale height, which is typically 1-2 km.
From this, the specific attenuation as a function of height can be approximately modeled as
\begin{align}
	\label{po3}
	h_{L,\text{dB/km}}^{o,w}(f_c,H) = h_{L,\text{dB/km}}^{o,w,0}(f_c) \exp\left(-H/H_\text{scale}\right).
\end{align}
In our system model, both CU and UD links are slant.
For a slant atmospheric path from height $H_1$ to $H_2$ at an angle $\psi$, the total atmospheric attenuation is obtained by integration from \eqref{po3} as
\begin{align}
	\label{po4}
	h_{L,\text{dB/km}}^{o,w}(f_c) \simeq 
	\frac{h_{L,\text{dB/km}}^{o,w,0}(f_c) \left( e^{-H_1/H_\text{scale}} - e^{-H_2/H_\text{scale}} \right) H_s}{\sin(\psi)}.
\end{align}
%


\subsection{3D Antenna Pattern}
%
%
Nowadays, advances in the fabrication of antenna array technology at mmWave bands allow the creation of large antenna arrays, with high antenna pattern gain in a cost effective and compact form, in order to compensate the negative effects of high propagation attenuation at mmWave frequencies. 
As mentioned, we consider a uniform array antenna, comprising $N_x\times N_y$ antenna elements in the $x$- and $y$-directions, where the space between antenna elements in the $x$- and $y$-directions are $d_{x}$ and $d_{y}$, respectively.
The array radiation gain is mainly formulated in the direction of $\theta$ and $\phi$. In our model, $\theta$ and $\phi$ can be defined as  functions of random variables (RVs) $\theta_{x}$ and $\theta_{y}$ as follows:
\begin{align}
	\label{f_2}
	\theta  &= \tan^{-1}\left(\sqrt{\tan^2(\theta_{x})+\tan^2(\theta_{y})}\right), \nonumber\\
	\phi    &=\tan^{-1}\left({\tan(\theta_{y})}\big/{\tan(\theta_{x})}\right).
\end{align} 
By taking into account the effect of all elements, the array radiation gain in the direction of angles $\theta_{x}$ and $\theta_{y}$ will be:
\begin{align}
	\label{p_1}
	G(\theta_{x},\theta_{y})  = G_0(N) \,
	\underbrace{G_e(\theta_{x},\theta_{y}) \,  G_a(\theta_{x},\theta_{y})}_{G'(\theta_{x},\theta_{y'})},
\end{align}
where $G_a$ is an array factor, $G_e$ is single element radiation pattern and $G_0$ is a constant defined in the sequel. 
From the 3GPP single element radiation pattern, $G_e = 10^{G_{e,\textrm{3dB}}/10}$ of each single antenna element is obtained as \cite{niu2015survey}
\begin{align}   
	\left\{
	\begin{array}{rl}
		&\!\!\!\!\!\!\! G_{e\textrm{3dB}} = G_{\textrm{max}} - \min\left\{-(G_{e\textrm{3dB,1}}+G_{e\textrm{3dB,2}}),F_m 
		\right\}, \nonumber \\
		&\!\!\!\!\!\!\! G_{e\textrm{3dB,1}} =  -\min \left\{ - 12\left(\frac{\theta_e-90}{\theta_{e\textrm{3dB}}}\right)^2,
		G_{\textrm{SL}}\right\},      
		\nonumber \\
		&\!\!\!\!\!\!\! G_{e\textrm{3dB,2}} = -\min \left\{ - 12\left(\frac{\theta_{x}}{\phi_{e\textrm{3dB}}}\right)^2,
		F_m\right\},
		\nonumber \\
		&\!\!\!\!\!\!\! \theta_e             = \tan^{-1}\left( \frac{\sqrt{1+\sin^2(\theta_{x})}}
		{\sin(\theta_{y'})} \right), 
	\end{array} \right. \nonumber
\end{align}
where $\theta_{e\textrm{3dB}}=65^{\circ}$ and $\phi_{e\textrm{3dB}}=65^{\circ}$ are the vertical and horizontal 3D beamwidths, respectively, $G_{\textrm{max}}=8$ dBi is the maximum directional gain of the antenna element, $F_m=30$ dB is the front-back ratio, and $G_{\textrm{SL}}=30$ dB is the side-lobe level limit. 

If the amplitude excitation of the entire array is uniform, then the array factor $G_a(\theta_{x},\theta_{y})$ for a square array of $N\times N$ elements can be obtained as \cite[eqs. (6.89) and (6.91)]{balanis2016antenna}
\begin{align}
	\label{f_1}
	G_a(\theta_{x},\theta_{y}) &= 
	\left( \frac{\sin\left(\frac{N (k d_{x} \sin(\theta)\cos(\phi)+\beta_{x})}{2}\right)} 
	{N\sin\left(\frac{k d_{x} \sin(\theta)\cos(\phi)+\beta_{x}}{2}\right)}
	\right. \nonumber \\
	&\times \left. \frac{\sin\left(\frac{N (k d_{y} \sin(\theta)\sin(\phi)+\beta_{y})}{2}\right)} 
	{N\sin\left(\frac{k d_{y} \sin(\theta)\sin(\phi)+\beta_{y}}{2}\right)}\right)^2,
\end{align}
where $\beta_{x}$ and $\beta_{y}$ are progressive phase shift between the elements along the  $x$ and $y$ axes, respectively. 
For a fair comparison between antennas with different $N$, we assume that the total radiated power of antennas with different $N$ are the same. From this, we have
\begin{align}
	\label{cv}
	G_0(N)=\left(   \int_0^{\pi}\int_0^{2\pi} G'(\theta,\phi) \sin(\theta) d\theta d\phi   \right)^{-1}.
\end{align}
More details on the elements and array radiation pattern are provided in \cite{niu2015survey,balanis2016antenna}.
In addition, and without loss of generality, it is assumed that $\beta_{x}=\beta_{y}=0$.
%


\section{Performance Analysis}
\subsection{Single UAV Relay}
For a given region with physical parameters such as air pressure, temperature, $\psi_{s,\text{min}}$, $\psi_{d,\text{min}}$, and $H_{sd}$, our aim is to adjust the tunable system parameters such as $N_{sx}$, $N_{sy}$, $N_{dx}$, $N_{dy}$, $N_{usx}$, $N_{usy}$, $N_{udx}$, $N_{udy}$, $H_u$ and $L_{sc}$, to improve system performance in terms of 
average capacity and the outage probability. These two metrics are very important in the design of wireless communication systems.
Our objective is  to maximize the channel capacity with the outage probability as a constraint (it is less than a threshold, i.e., $\mathbb{P}_\text{out}<\mathbb{P}_\text{out,th}$, where $\mathbb{P}_\text{out,th}$ is determined based on the requested QoS).
Our optimization problem is formulated as:
\begin{subequations} \label{opt1} 
	\begin{IEEEeqnarray}{l} 
		\displaystyle \max_{ \substack{  
				N_{sx}, N_{sy}, N_{dx}, N_{dy}, \\
				N_{usx}, N_{usy}, N_{udx}, N_{udy}, \\ H_u, L_{su}     
		}  }  ~~~~~~~{\bar{\mathbb{C}}_{e2e}}\\
		~~~~~~~~~~~\textrm{s.t.}  ~~~~~~~~~~~~\mathbb{P}_\text{out}< \mathbb{P}_\text{out,tr}~~~~~~~~~ \label{e3}\\
		~~~~~~~~~~~~~~~~~ ~~~~~~~~~H_u > H_{u,\text{min}}, \label{e2}
	\end{IEEEeqnarray}
\end{subequations}
where $\bar{\mathbb{C}}_{e2e}$ is the average channel capacity during the UAV flight time.
%
Constraint  \eqref{e2} is used to guarantee that the UAV is in the LoS of both CN and RA throughout the entire flight path. Therefore, the minimum height of the UAV should be
\begin{align}
	\label{ds1}
	&H_{u,\text{min}}= \\
	&\max\left\{(L_{sc}+\frac{L_{u1}}{2}) \sin(\psi_{s,\text{min}}),~(L_{dc}+\frac{L_{u1}}{2}) \sin(\psi_{d,\text{min}})\right\}, \nonumber
\end{align}
to ensure it satisfies the LoS for both links.
In \eqref{ds1}, we have
$L_{sc}=\sqrt{L_{s,\text{min}}^2-H_u^2}+\frac{L_{u1}}{2}$, 
$L_{dc}=\sqrt{L_{d,\text{min}}^2-H_u^2}+\frac{L_{u1}}{2}$, where $L_{s,\text{min}}$ is the link length between CN and $B_{p1}$, while $L_{d,\text{min}}$ is the link length between RA and $B_{p2}$.
We consider that the points $B_{p}$, $B_{p1}$, and $B_{p2}$ are in $[x,y]=[0,0]$, $[\frac{L_{u1}}{2},0]$, and  $[-\frac{L_{u1}}{2},0]$, respectively.
Let $\mathcal{R}_1$ indicates the path of a semicircle that starts from point $B_{p1}$ and reaches point $B_{p2}$. 
Therefore, each point on $\mathcal{R}_1$ in the $[x-y]$ plane is specified as follows
\begin{align}
	x_{u} = \frac{L_{u1}}{2}\cos(\theta_{R1}), ~~~
	y_{u} = \frac{L_{u1}}{2}\sin(\theta_{R1}), 
\end{align}
where $0<\theta_{R1}<\pi$.
From this, the average channel capacity can be formulated as
\begin{align}
	\label{df1}
	\bar{\mathbb{C}}_{e2e} = \frac{1}{\pi}\int_{\theta_{R1}=0}^\pi \mathbb{C}_{e2e|\theta_{R1}}  \text{d}\theta_{R1},
\end{align}
where $\mathbb{C}_{e2e|\theta_{R1}}$ is the average end-to-end channel capacity conditioned on $\theta_{R1}$. For our system model, $\mathbb{C}_{e2e|\theta_{R1}}$ is a function of random variables (RVs) 
$\theta_{sx}$, $\theta_{sy}$, $\theta_{usx}$, $\theta_{usy}$,
$\theta_{dx}$, $\theta_{dy}$, $\theta_{udx}$, and $\theta_{udy}$,  and is obtained in \eqref{c7}.
\begin{figure*}[!t]
	\normalsize
	\begin{align}
		\label{c7}
		&\mathbb{C}_{e2e|\theta_{R1}} = 
		\frac{1}{16\pi^4 
			\sigma_{sx} \sigma_{sy} \sigma_{rsx} \sigma_{rsy} 
			\sigma_{dx} \sigma_{dy} \sigma_{rdx} \sigma_{rdy} }
		\int_0^{\pi/2} \int_0^{\pi/2} \int_0^{\pi/2} \int_0^{\pi/2} 
		\int_0^{\pi/2} \int_0^{\pi/2} \int_0^{\pi/2} \int_0^{\pi/2}
		\mathbb{C}'_{e2e|\theta_{R1}}  \nonumber \\
		& \times  
		\exp\left( -\frac{(\theta_{sx}-\mu_{sx})^2}{2\sigma_{sx}}  \right)
		\exp\left( -\frac{(\theta_{sy}-\mu_{sy})^2}{2\sigma_{sy}}  \right)
		\exp\left( -\frac{(\theta_{usx}-\mu_{usx})^2}{2\sigma_{usx}}  \right)
		\exp\left( -\frac{(\theta_{usy}-\mu_{usy})^2}{2\sigma_{usy}}  \right)
		\nonumber \\
		& \times  
		\exp\left( -\frac{(\theta_{dx}-\mu_{dx})^2}{2\sigma_{dx}}  \right)
		\exp\left( -\frac{(\theta_{dy}-\mu_{dy})^2}{2\sigma_{dy}}  \right)
		\exp\left( -\frac{(\theta_{udx}-\mu_{udx})^2}{2\sigma_{udx}}  \right)
		\exp\left( -\frac{(\theta_{udy}-\mu_{udy})^2}{2\sigma_{udy}}  \right)
		\nonumber \\
		&\times 
		\text{d}\theta_{sx}  \text{d}\theta_{sy} \text{d}\theta_{usx}  \text{d}\theta_{usy}
		\text{d}\theta_{dx}  \text{d}\theta_{dy} \text{d}\theta_{udx}  \text{d}\theta_{udy},
	\end{align}
	\textrm{where}
	\label{c8}
	\begin{align}
		\mathbb{C}'_{e2e|\theta_{R1}} &= \text{min}\Big\{   
		\log_2\big(1 + \Gamma_{s}(\theta_{sx},\theta_{sy},\theta_{usx},\theta_{usy}|\theta_{R1} ) \big), ~
		\log_2\big(1 + \Gamma_{d}(\theta_{dx},\theta_{dy},\theta_{udx},\theta_{udy}|\theta_{R1} ) \big)\Big\}  ~~~~~~~~~~\nonumber\\
		&=\log_2\Big(1 + \text{min}  \Big\{ 
		\Gamma_{s}(\theta_{sx},\theta_{sy},\theta_{usx},\theta_{usy}|\theta_{R1} ),~ 
		\Gamma_{d}(\theta_{dx},\theta_{dy},\theta_{udx},\theta_{udy}|\theta_{R1} )  \Big\}
		\Big),
	\end{align}
    \begin{align}
    	\label{ro}
    	&\Gamma_{q}(\theta_{qx},\theta_{qy},\theta_{uqx},\theta_{uqy}|\theta_{R1} ) = \frac{P_{t,q} h_L(L_q(\theta_{R1},L_{sc},H_u)) }{\sigma_n^2}  
    	 G_0(N_{qx},N_{qy}) G_0(N_{uqx},N_{uqy}) G_e(\theta_{qx},\theta_{qy}) G_e(\theta_{uqx},\theta_{uqy}) \nonumber  \\
    	& \times
    	\left( \frac{\sin\left(\frac{N_{qx} (k d_{qx} 
    			\sin\left(\theta_{qxy}\right)
    			\cos\left( \tan^{-1}\left(\frac{\tan(\theta_{qy})}{\tan(\theta_{qx})}\right) \right)+\beta_{qx})}{2}\right)} 
    	{N_{qx}\sin\left(\frac{k d_{qx} 
    			\sin\left( \theta_{qxy} \right)
    			\cos\left( \tan^{-1}\left(\frac{\tan(\theta_{qy})}{\tan(\theta_{qx})}\right) \right)+\beta_{qx}}{2}\right)}
    	\frac{\sin\left(\frac{N_{qy} (k d_{qy} 
    			\sin\left( \theta_{qxy} \right)
    			\sin\left( \tan^{-1}\left(\frac{\tan(\theta_{qy})}{\tan(\theta_{qx})}\right) \right)+\beta_{qy})}{2}\right)} 
    	{N_{qy}\sin\left(\frac{k d_{qy} \sin\left(\theta_{qxy} \right)
    			\sin\left( \tan^{-1}\left(\frac{\tan(\theta_{qy})}{\tan(\theta_{qx})}\right) \right)+\beta_{qy}}{2}\right)}\right)^2	 \nonumber \\	
    	&\times \left( \frac{\sin\left(\frac{N_{uqx} (k d_{uqx} 
    			\sin\left(\theta_{uqxy}\right)
    			\cos\left( \tan^{-1}\left(\frac{\tan(\theta_{uqy})}{\tan(\theta_{uqx})}\right) \right)+\beta_{uqx})}{2}\right)} 
    	{N_{uqx}\sin\left(\frac{k d_{uqx} \sin\left(\theta_{uqxy}\right)
    			\cos\left( \tan^{-1}\left(\frac{\tan(\theta_{uqy})}{\tan(\theta_{uqx})}\right) \right)+\beta_{uqx}}{2}\right)}
    	\frac{\sin\left(\frac{N_{uqy} (k d_{uqy} 
    			\sin\left(\theta_{uqxy}\right)
    			\sin\left( \tan^{-1}\left(\frac{\tan(\theta_{uqy})}{\tan(\theta_{uqx})}\right) \right)+\beta_{uqy})}{2}\right)} 
    	{N_{uqy}\sin\left(\frac{k d_{uqy} \sin\left(\theta_{uqxy}\right)
    			\sin\left( \tan^{-1}\left(\frac{\tan(\theta_{uqy})}{\tan(\theta_{uqx})}\right) \right)+\beta_{uqy}}{2}\right)}\right)^2.
    \end{align}
	\hrulefill
\end{figure*}
In \eqref{ro}, we have
$\theta_{qxy}=\tan^{-1}\left(\sqrt{\tan^2(\theta_{qx})+\tan^2(\theta_{qy})}\right)$, and
$\theta_{uqxy}=\tan^{-1}\left(\sqrt{\tan^2(\theta_{uqx})+\tan^2(\theta_{uqy})}\right)$.
Moreover, $L_s$ and $L_d$ are functions of $H_u$, $L_{sc}$, and $\theta_{R1}$ as
\begin{align}\label{r1}
	& L_s \!\!=\! \sqrt{ \! \left(\!L_{sc}-\frac{L_{u1}}{2}\cos(\theta_{R1})\!\right)^2+\frac{L_{u1}^2}{4}\sin^2(\theta_{R1}) + H_u^2   }\nonumber \\
	& L_d \!\!=\! \sqrt{ \! \left(\!L_{dc}\!+\!\frac{L_{u1}}{2}\cos(\theta_{R1})\!\right)^2\!+\frac{L_{u1}^2}{4}\sin^2(\theta_{R1}) + H_u^2   }.	
\end{align}
As can be seen, calculating the channel capacity from Eqs. \eqref{c7}-\eqref{ro} requires solving a 9-dimensional integral equation numerically, which is very time consuming.
In order to analyze and design the system parameters optimally, it is necessary to provide more tractable and well-formed analytical expressions for performance metrics as a function of channel parameters. Therefore, in the following, we first present the distribution of end-to-end SNR and then use it to calculate the performance metrics such as outage probability and channel capacity. 

{\bf Theorem 1.}
{\it The distribution of end-to-end  SNR conditioned on $\theta_{R1}$ is derived as}
\begin{align}
	\label{sm3}
	&f_{\Gamma_{q}|\theta_{R1}}(\Gamma_{q}|\theta_{R1}) = 
	\sum_{j_q =1}^{KJ_q} \sum_{j_u =1}^{KJ_u}  \mathbb{T}_{q}(j_q,N_{qx})  \mathbb{T}_{uq}(j_u,N_{uqx})   
	\nonumber \\
	&~~~~~~~~~~~~~~~~~~~~~~\times \delta\left( \Gamma_{q} - \Gamma'_{q}(j_q,j_u) \right),  
\end{align}
{\it where $\delta(\cdot)$ is Dirac delta function, $\mathbb{T}_{q}(j_q,N_{qx})$ is derived in \eqref{sm6}, 
	$\Gamma'_{q}(j_q,j_u) =\Gamma''_{q}(j_q,j_u)h_L(L_q(\theta_{R1},L_{sc},H_u)) $, and}
\begin{align}
	\label{sm4}
	&\Gamma''_{q}(j_q,j_u) = \frac{ 10^{G_\text{max}/5} G_0(N_{qx},N_{qy}) G_0(N_{uqx},N_{uqy}) P_{t,q} }{N_{qx}^2 N_{uqx}^2 \sigma_n^2}  
	\nonumber \\
	&       
	\left( \frac{\sin\left(\frac{N_{qx} k d_{qx} 
			\sin\left(\frac{2j_q}{J_q N_{qx}}\right)
		}{2}\right)} 
	{\sin\left(\frac{k d_{qx} 
			\sin\left( \frac{2j_q}{J_q N_{qx}} \right)
		}{2}\right)} 
	\frac{\sin\left(\frac{N_{uqx} k d_{uqx} 
			\sin\left(\frac{2j_u}{J_u N_{uqx}}\right)
		}{2}\right)} 
	{\sin\left(\frac{k d_{uqx} \sin\left(\frac{2j_u}{J_u N_{uqx}}\right)
		}{2}\right)}
	\right)^2.
\end{align}
{\it Also, $\mathbb{T}_{uq}(j_u,N_{uqx})$ is obtained from \eqref{sm6} by substituting $j_u$, $J_u$, $m_u$, $M_u$, $N_{uqw}$, $\mu_{uqx}$, $\mu_{uqy}$, $\sigma_{uqx}$, and  $\sigma_{uqy}$ instead of $j_q$, $J_q$, $m_q$, $M_q$, $N_{qw}$, $\mu_{qx}$, $\mu_{qy}$, $\sigma_{qx}$, and  $\sigma_{qy}$, respectively.}
%
\begin{figure*}[!t]
	\normalsize
	\begin{align}  
		\label{sm6}
		&\mathbb{T}_{q}(j_q,N_{qx}) = \sum_{m_q=1}^{M_q}
		\left[
		Q\left(-\sqrt{\frac{4 j_q^2}{J_q^2 N_{qx}^2\sigma_{qx}^2}-\left(\frac{2 j_q(m_q-1)}{J_q N_{qx}M_q \sigma_{qx}}\right)^2}-\frac{\mu_{qx}}{\sigma_{qx}}
		\right) 
		- Q\left(\sqrt{\frac{4 j_q^2}{J_q^2 N_{qx}^2\sigma_{qx}^2}-\left(\frac{2 j_q(m_q-1)}{J_q N_{qx}M_q \sigma_{qx}}\right)^2}-\frac{\mu_{qx}}{\sigma_{qx}}
		\right)  
		\right]  \nonumber \\
		&\times \left[
		Q\left(  {\frac{2 j_q(m_q-1)}{J_q N_{qx}M_q\sigma_{qy}}-\frac{\mu_{qy}}{\sigma_{qy}}}  \right)  -   
		Q\left({\frac{2 j_q m_q}{J_q N_{qx}M_q\sigma_{qy}}-\frac{\mu_{qy}}{\sigma_{qy}}  }  \right) 
		+
		Q\left({-\frac{2 j_q m_q}{J_q N_{qx}M_q\sigma_{qy}}-\frac{\mu_{qy}}{\sigma_{qy}}  }  \right)  -
		Q\left({-\frac{2 j_q(m_q-1)}{J_q N_{qx}M_q\sigma_{qy}}-\frac{\mu_{qy}}{\sigma_{qy}}  }  \right)
		\right] \nonumber \\
		&-\left[
		Q\left(-\sqrt{\frac{4 (j_q-1)^2}{J_q^2 N_{qx}^2\sigma_{qx}^2}-\left(\frac{2 (j_q-1)(m_q-1)}{J_q N_{qx}M_q \sigma_{qx}}\right)^2}-\frac{\mu_{qx}}{\sigma_{qx}}
		\right) 
		- Q\left(\sqrt{\frac{4 (j_q-1)^2}{J_q^2 N_{qx}^2\sigma_{qx}^2}-\left(\frac{2 (j_q-1)(m_q-1)}{J_q N_{qx}M_q \sigma_{qx}}\right)^2}-\frac{\mu_{qx}}{\sigma_{qx}}
		\right)  
		\right]  \nonumber \\
		&\times \left[
		Q\left(  {\frac{2 (j_q-1)(m_q-1)}{J_q N_{qx}M_q\sigma_{qy}}-\frac{\mu_{qy}}{\sigma_{qy}}}  \right)  -   
		Q\left({\frac{2 (j_q-1) m_q}{J_q N_{qx}M_q\sigma_{qy}}-\frac{\mu_{qy}}{\sigma_{qy}}  }  \right) 
		+
		Q\left({-\frac{2 (j_q-1) m_q}{J_q N_{qx}M_q\sigma_{qy}}-\frac{\mu_{qy}}{\sigma_{qy}}  }  \right)  \right. \nonumber \\
		&-\left. Q\left({-\frac{2 (j_q-1)(m_q-1)}{J_q N_{qx}M_q\sigma_{qy}}-\frac{\mu_{qy}}{\sigma_{qy}}  }  \right)
		\right] 
	\end{align}
	\hrulefill
\end{figure*}
%

\begin{IEEEproof}
	Please refer to Appendix \ref{AppA}.
\end{IEEEproof}
As can be seen, the closed-form expression provided in \eqref{sm3} is very simple and calculates the distribution of end-to-end SNR conditioned on $\theta_{R1}$ based on the sum of a series of  Dirac delta functions.
The Dirac delta function is due to the approximation of the antenna pattern with $K J_q$ and $K J_u$ sectors. 
As the number of sectors increases, we expect the sectorized antenna pattern to approach the actual pattern.
In the following, another approximate model for the end-to-end SNR distribution is presented with a little less accuracy than \eqref{sm3}, but with a lower computational volume.

{\bf Proposition 1.}
{\it The distribution of the end-to-end  SNR conditioned on $\theta_{R1}$ given in \eqref{sm3} can be simplified as}
\begin{align}
	\label{sm7}
	&f_{\Gamma_{q}|\theta_{R1}}(\Gamma_{q}|\theta_{R1}) = 
	\sum_{j_q =1}^{KJ_q} \sum_{j_u =1}^{KJ_u}       
	\mathbb{B}_{q}(j_q,N_{qx})
	\delta\Big( \Gamma_{q} - \Gamma'_{q}(j_q,j_u) \Big),  
\end{align}
{\it with}
\begin{align}  
	\label{sm8}
	&\mathbb{B}_{q}(j_q,N_{qx}) = 
	\exp\left(-\frac{2 (j_q-1)^2}{J_q^2 N_{qx}^2\sigma_{qc}^2} \right)
	-\exp\left(-\frac{2 j_q^2}{J_q^2 N_{qx}^2\sigma_{qc}^2} \right),
\end{align} 
{\it where $\sigma_{qc}^2$ is obtained in \eqref{k3}.}
\begin{IEEEproof}
	Please refer to Appendix \ref{AppB}.
\end{IEEEproof}

As can be seen, \eqref{sm7} has a lower computational load than \eqref{sm3}, but, its accuracy is lower. In the following, the accuracy of the closed-form expressions with the results obtained from Monte-Carlo simulations is examined. However, the main feature of the provided analytical expression is that it is tractable, and will allow us to properly calculate the closed-form expressions for outage probability and channel capacity.

{\bf Proposition 2.}
{\it Based on the channel distribution provided in Theorem 1, the average end-to-end channel capacity is obtained as}
\begin{align}
	\label{sn7}
	\bar{\mathbb{C}}_{e2e} = \frac{1}{\pi}\int_{\theta_{R1}=0}^\pi \min\left\{
	\mathbb{C}_{su|\theta_{R1}},
	\mathbb{C}_{du|\theta_{R1}}
	\right\}  \text{d}\theta_{R1}, 
\end{align}
{\it where}
\begin{align}  
	\label{sn8}
	&\mathbb{C}_{qu|\theta_{R1}} = 
	\sum_{j_q =1}^{KJ_q} \sum_{j_u =1}^{KJ_u}                \mathbb{T}_{q}(j_q,N_{qx})  \mathbb{T}_{uq}(j_u,N_{uqx}) \nonumber \\ 
	&~~~~~~~~\times \log_2\Big(1 + \Gamma''_{q}(j_q,j_u)h_L(L_q(\theta_{R1},L_{sc},H_u)) \Big).
\end{align} 
{\it Also, based on the channel distribution provided in Proposition 1, another closed-form expression for $\mathbb{C}_{qu|\theta_{R1}}$ with lower computational load is obtained as}
\begin{align}  
	\label{sn9}
	&\mathbb{C}_{qu|\theta_{R1}} = 
	\sum_{j_q =1}^{KJ_q} \sum_{j_u =1}^{KJ_u}                \mathbb{B}_{q}(j_q,N_{qx})  
	\log_2\Big(1 + \Gamma'_{q}(j_q,j_u) \Big).
\end{align} 
\begin{IEEEproof}
	Please refer to Appendix \ref{AppC}.
\end{IEEEproof}

{\bf Proposition 3.}
{\it Based on the channel distribution provided in Theorem 1, the end-to-end outage probability of the considered system conditioned on $\theta_{R1}$ is derived as}
\begin{align}
	\label{gm1}
	\mathbb{P}_{\text{out}|\theta_{R1}} &= \mathbb{P}_{\text{out,su}|\theta_{R1}} + \mathbb{P}_{\text{out,du}|\theta_{R1}} - \mathbb{P}_{\text{out,su}|\theta_{R1}}\mathbb{P}_{\text{out,du}|\theta_{R1}}, 
\end{align}
{\it where}
\begin{align}
	\label{gm2}
	&\mathbb{P}_{\text{out,qu}|\theta_{R1}} = 
	\sum_{j_q =1}^{KJ_q} \sum_{j_u =1}^{KJ_u}  \mathbb{T}_{q}(j_q,N_{qx})  \mathbb{T}_{uq}(j_u,N_{uqx})   
	\nonumber \\
	&~~~~~~~~~~~~~~~~~~~~~~\times \mathbb{Y}\left( \Gamma_\text{th} - \Gamma'_{q}(j_q,j_u) \right).  
\end{align}
{\it Also, based on the channel distribution provided in Proposition 1, another closed-form expression for $\mathbb{P}_{\text{out,qu}|\theta_{R1}}$ with lower computational load is obtained as}
\begin{align}
	\label{gm3}
	&\mathbb{P}_{\text{out,qu}|\theta_{R1}} = 
	\sum_{j_q =1}^{KJ_q} \sum_{j_u =1}^{KJ_u}  \mathbb{B}_{q}(j_q,N_{qx})  
	\mathbb{Y}\left( \Gamma_\text{th} - \Gamma'_{q}(j_q,j_u) \right).  
\end{align}
\begin{IEEEproof}
	Please refer to Appendix \ref{AppD}.
\end{IEEEproof}

The main important point about the closed-form expressions presented in Propositions 2 and 3 is that in addition to being tractable and very well formed, they are a function of all key channel parameters and we can easily analyze the effect of channel parameters on the outage probability and channel capacity of the considered system with a greater speed and lower computational load. 

\subsection{Multiple Relay System}
%
Although the design of a multi-relay system is slightly more complicated than a single-relay system, the expressions obtained for a single-relay system can be easily extended to a multi-relay system as follows. 

{\bf Proposition 4.}
{\it  The end-to-end  channel capacity  and outage probability of an $M$ relay system is derived respectively as}
\begin{align}
	\label{g1}
	\bar{\mathbb{C}}_{e2e} \simeq \frac{1}{\pi}\int_{\theta_{R1}=0}^\pi \min\left\{
	\mathbb{C}_{su|\theta_{R1}},
	\mathbb{C}_{du|\theta_{R1}}, \mathbb{C}_{uu}
	\right\}  \text{d}\theta_{R1}, 
\end{align}
{\it and}
\begin{align}
	\label{g2}
	\mathbb{P}_{\text{out}|\theta_{R1}} &= 1-(1-\mathbb{P}_{\text{out,su}|\theta_{R1}})(1-\mathbb{P}_{\text{out,du}|\theta_{R1}}) \nonumber \\
	&~~~\times(1-\mathbb{P}_{\text{out,uu}|\theta_{R1}})^{M-1},
\end{align}
{\it where $\mathbb{C}_{uu}$ and $\mathbb{P}_{\text{out,uu}|\theta_{R1}}$ are respectively obtained from Eqs. \eqref{sn7} and \eqref{gm3} by substituting parameters 
$N_{tx1}$, $N_{ty1}$, $N_{rx2}$, $N_{ry2}$, $\sigma_{uw1}$, $\sigma_{uw2}$, $\mu_{uw1}$, $\mu_{uw2}$, $L_{uu}$, $J_u$, and $h_L(L_{uu},H_u)$ instead of 
$N_{qx}$, $N_{qy}$, $N_{uqx}$, $N_{uqy}$,   $\sigma_{qw}$, $\sigma_{uqw}$, $\mu_{qw}$, $\mu_{uqw}$, $L_q$, $J_q$, and $h_L(L_q(\theta_{R1},L_{sc},H_u))$, respectively.}
\begin{IEEEproof}
	Based on the results of next remark and by following the method adopted in Appendices \ref{AppA}, \ref{AppB}, and \ref{AppD}, the results will be proven. 
\end{IEEEproof}

As discussed, for a multi-relay system, it is assumed that the UAVs are rotating at the same speed and same angle $\theta_{R1}$ relative to each other. Therefore, along the entire circular flight path of the UAVs, the link length between the UAVs remains constant. Based on this, we can conclude the following remark. 

{\bf Remark 1.} {\it The inter UAV links are symmetric, and the optimal values for the parameters of an inter UAV link can be used for the rest of the UAV links, and as a result, the design of a multi-relay system will be quite similar to a two-relay system.}

\section{Simulations and Optimal System Design}
By providing comprehensive simulations, the performance of the single-relay as well as the multi-relay systems is examined. The values of the parameters used in the simulations are listed in Table \ref{I2}. The Monte-Carlo simulations also are used to show the accuracy of the provided analytical expressions. 
In the following, the single-relay system will be examined first, and then the multi-relay system will be studied for longer link lengths.

\begin{table}
	\caption{Parameter values for simulations of single relay system.} 
	\centering 
	\begin{tabular}{| l |c || l| c|} 
		\hline\hline 
		{\bf Parameters} & {\bf Values} &
		{\bf Parameters} & {\bf Values}  \\ [.5ex] 
		\hline\hline 
		%
		$P_{t,s} $ & 1 W &
		$P_{t,d} $ & 200 mW\\ \hline
		$N_{qw}$ & 12-18 &
		$N_{uqw}$ & 6-18 \\ \hline
		$f_c$ &  70 GHz &
		%
			$\mathbb{P}_\text{out,tr}$ & $10^{-3}$\\ \hline
			$\rho_0$&   $7.5 ~\text{g/m}^3$  &
			$T$ & $20^o$C\\ \hline
			$\beta_{qw}=\beta_{uqw}$ & 0 &
			$L_{u1}$ & 3.5 km \\ \hline
			$H_\text{scale}$ &  1.5 km &  $L_{sd}$ & 17 km \\ \hline
			$\psi_{d,\text{min}}$	& $15^o$  & $\psi_{s,\text{min}}$
			& $10^o$\\ \hline
			$d_{qw}=d_{uqw}$ & $\lambda/2$  & $\sigma_{uqx}\&\sigma_{uqy}$  & $2^o\&0.5$ \\ \hline
			$\sigma_{qw}\&\mu_{qw}$ &  $0.5^o \& 0.3^o$  & $\mu_{uqx}\&\mu_{uqy}$  & $0.8^o \& 0.2^o$ \\ 
			\hline \hline              
		\end{tabular}
		\label{I2} 
	\end{table}

\subsection{Single Relay Case}	
\begin{figure}
	\centering
	\subfloat[] {\includegraphics[width=3 in]{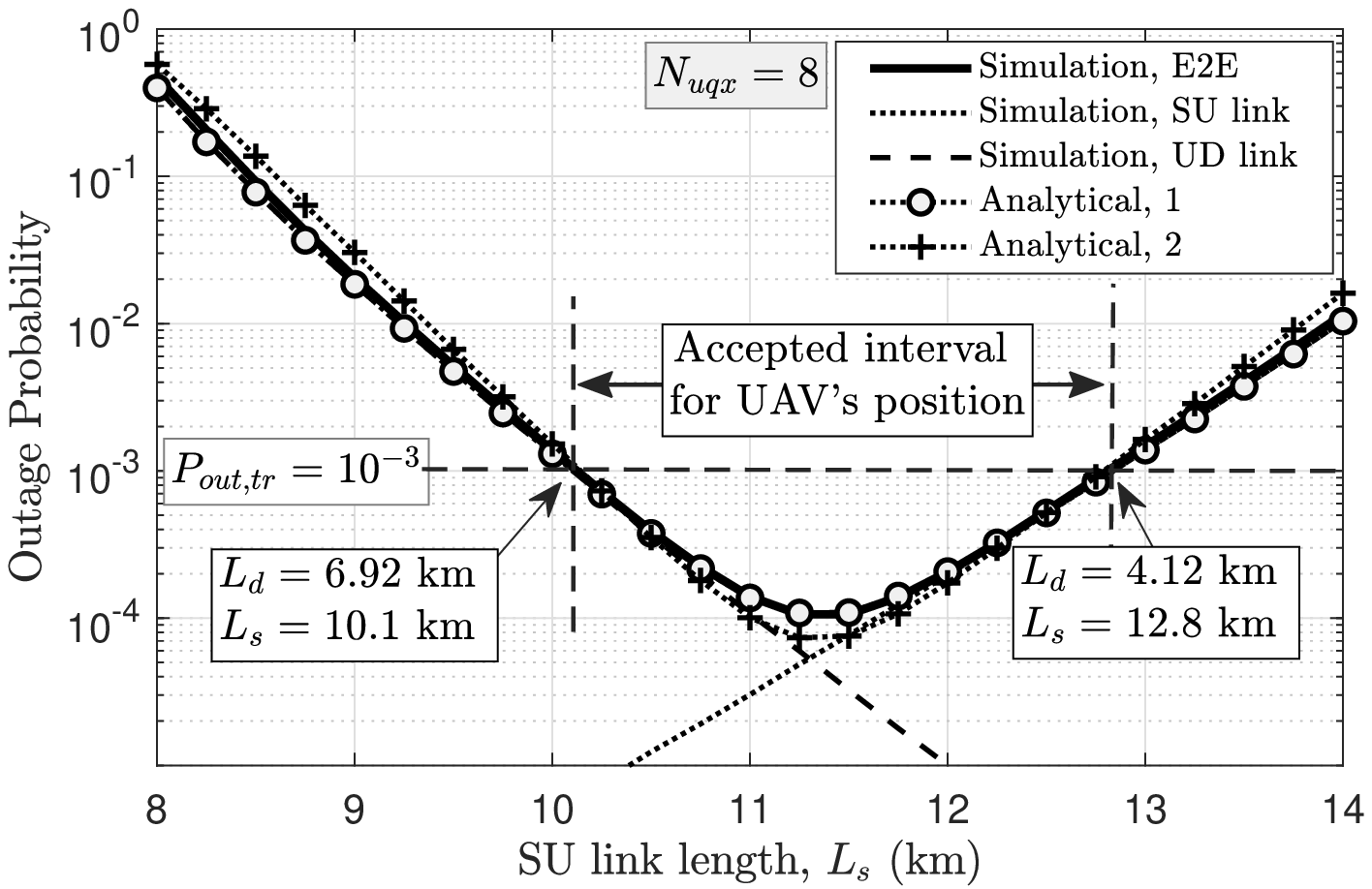}
		\label{xc1}
	}
	\hfill
	\subfloat[] {\includegraphics[width=3 in]{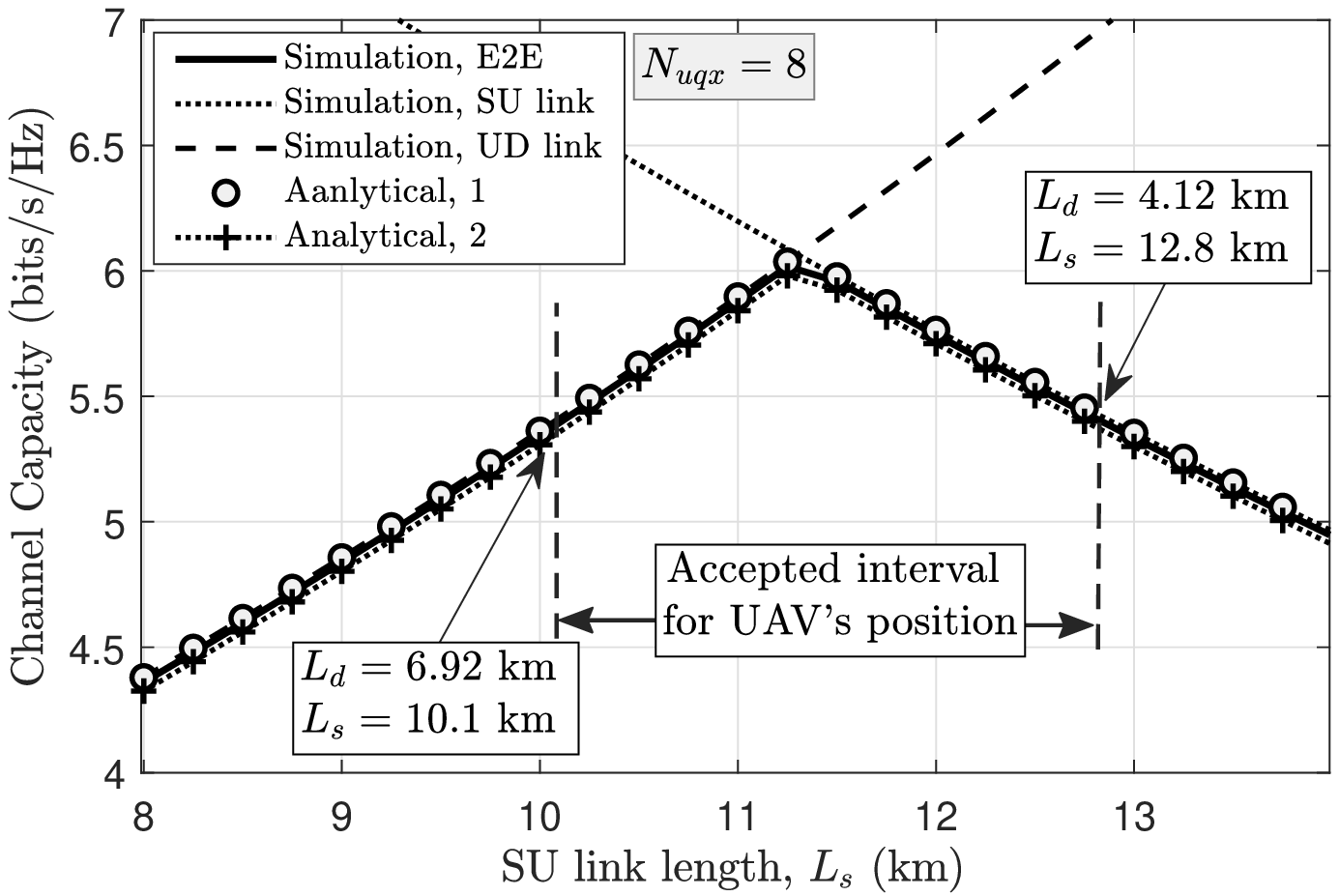}
		\label{xc2}
	}
	\caption{E2E performance of single-relay system versus $L_s$ and comparison with the performance of CU and UD links in terms of (a) outage probability and (b) channel capacity. }
	\label{xc3}
\end{figure}

For single relay systems, one of the important parameters is the optimal position for point $B_p$, which determines the average position of the UAV in a circular motion. As discussed, the location of the point $B_p$ is adjusted in sky with the parameter $L_{sc}$. Any change in the parameters $B_p$ and $L_{sc}$ affects the values of $L_s$ and $L_d$. 
In Fig. \ref{xc3}, the end-to-end outage probability and channel capacity are plotted versus $L_s$ for $N_{uqx}=12$, and $N_{uqy}=N_{qw}=N_\text{max}$. 
	%
As discussed in the previous section, the E2E performance depends on the performance of CU and UD links. 
Therefore, in Fig. \ref{xc3}, to get a better view, the performance of CU and UD links is also provided versus $L_s$.	
The results obtained from Fig. \ref{xc3} can be expressed in the following two remarks. 
	
{\bf Remark 2.} {\it For shorter links of $L_s$, the E2E performance can be well approximated with the performance of UD link. 
However, for longer links of $L_s$, E2E system performance is limited to the performance of SU link.
}
	
	{\bf Remark 3.} {\it The optimal value for $L_{sc}$ is very close to the length of $L_s$ for which the capacity of the SU link  is equal to the capacity of the UD link.}
	
	To justify {\bf Remark 2}, note that  by increasing $L_s$, the performance of the CU link decreases and at the same time $L_d$ decreases and consequently the performance of the UD link improves. 
	The accepted interval for $L_s$ and $L_d$ shown in Fig. \ref{xc1} is to guarantee condition \eqref{e3}. Based on \eqref{e3} and {\bf Remark 1}, we can conclude the following remark. 
	
	{\bf Remark 4.} {\it In order to guarantee constraint \eqref{e3} along the circle flight path, it is necessary that $L_s<L_{s,\text{max}}$ and $L_d<L_{d,\text{max}}$ where $L_{s,\text{max}}$ and $L_{d,\text{max}}$ are obtained as
		\begin{align}
			\label{e5}
			&\mathbb{P}_\text{out,tr} = \text{Prob}\left\{ 
			\Gamma_{s}(\theta_{sx},\theta_{sy},\theta_{usx},\theta_{usy}|L_{s,\text{max}} ) 
			<\Gamma_\text{th}\right\}, \\
			&\mathbb{P}_\text{out,tr} = \text{Prob}\left\{ 
			\Gamma_{d}(\theta_{dx},\theta_{dy},\theta_{udx},\theta_{udy}|L_{d,\text{max}} ) 
			<\Gamma_\text{th}\right\}.
		\end{align}
	}

Note that the use of Monte-Carlo simulations is very time consuming, especially for the lower values of outage probability. 
If Monte Carlo simulation is used to find the optimal values for tunable system parameters, it is necessary to independently run Monte-Carlo simulation in large numbers, and finally, select the optimal parameters from all the independent runs. 
As we see in the sequel, the performance of the considered system is highly dependent on the optimal values for the adjustable parameters such as the antenna patterns as well as the optimal position of the UAVs in 3D space. Therefore, for our system model, the search space to run Monte-Carlo simulation independently will be very large and thus, it will take a lot of time to optimally design the system parameters.
For this aim, in this work, the closed-form expressions for outage probability as well as channel capacity were presented as a function of all key parameters of the considered system, which have a much shorter run time than Monte-Carlo simulations.
In Figs. \ref{xc1} and \ref{xc2}, along with the simulation results, the two analytical expressions provided in Eqs. \eqref{gm2} and \eqref{gm3} for the outage probability as well as the two analytical expressions provided in Eqs. \eqref{sn8} and \eqref{sn9} for the channel capacity are plotted. The simulation results confirm the accuracy of the provided analytical expressions.
%

\begin{figure}
	\centering
	\subfloat[] {\includegraphics[width=3 in]{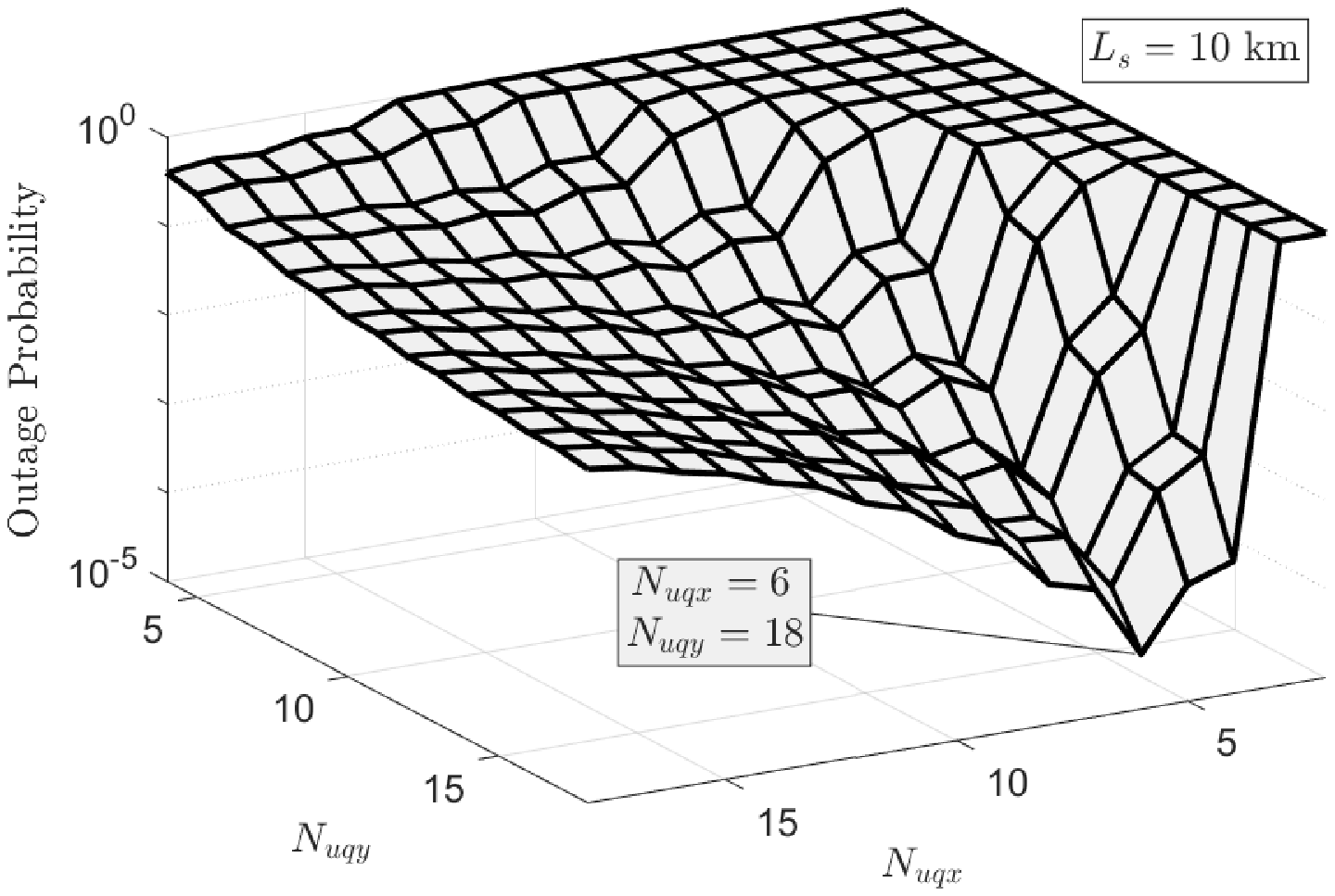}
		\label{xl1}
	}
	\hfill
	\subfloat[] {\includegraphics[width=3 in]{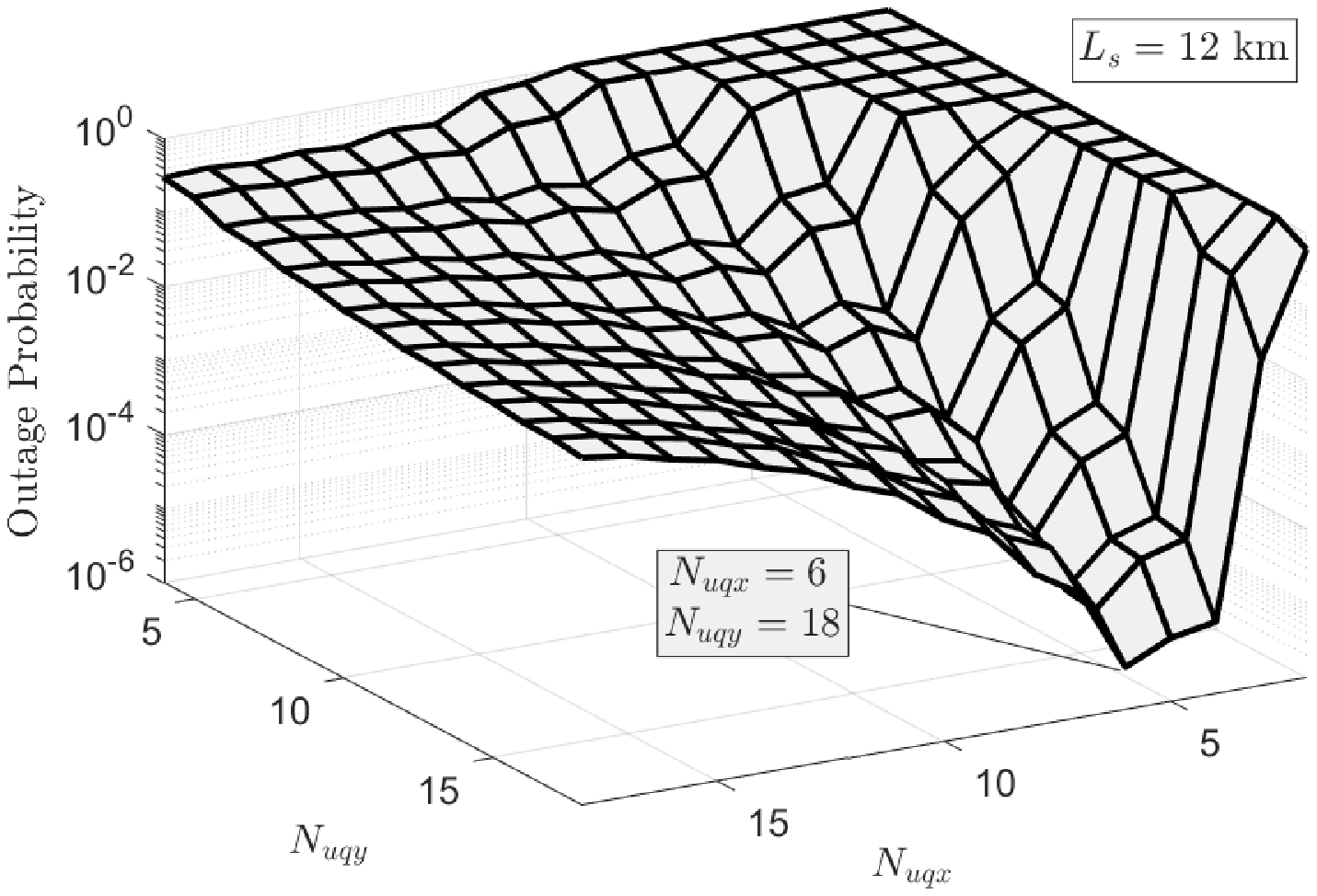}
		\label{xl2}
	}
	\caption{E2E outage probability of the single relay system versus joint $N_{uqx}$ and $N_{uqy}$ for two different values of (a) $L_s=10$ km and (b) $L_s=12$ km.}
	\label{xl3}
\end{figure}

%
In the direction of the UAV movement, the misalignment severity of antennas mounted on the fixed-wing UAV are higher than the misalignment severity in a direction perpendicular to the UAV movement.
Therefore, for the considered fixed-wing UAV, we have $\sigma_{uqx}>\sigma_{uqy}$, and thus, we expect the optimal number for antenna elements to be different along the $x_q$ and $y_q$ axes.
In Fig. \ref{xl3}, the outage probability of the single relay system is plotted versus both $N_{uqx}$ and $N_{uqy}$ for two different values of $L_s=10$ and 12 km.
The optimal selection of the antenna pattern in the direction of the $x_q$ axis, which has a larger angel-of-arrival standard deviation, is of higher importance than the antenna pattern in the direction of the $y_q$ axis.
In the $y_q$ axis, due to the lower $\sigma_{uqy}$, outage probability is more resistant to increasing the antenna gain pattern. As a result, with increasing $N_{uqy}$, the SNR in the receiver increases and thus, the performance of the system improves. Therefore, based on the results of Figs. \ref{xl3} and for both $L_s=10$ and 12 km, the optimal value for $N_{uqy}=N_\text{u,max}$ is 18.\footnote{In practice, due to the weight and aerodynamic limitations of the UAV payload, a very large antenna can not be used and we have to consider a maximum for $N_{quw}$.} 
However, in the direction of the $x_q$ axis, although the SNR increases by increasing $N_{uqx}$, for larger values of $N_{uqx}$, the beam width decreases and the system becomes more sensitive to misalignment errors and therefore, the outage probability increases. 
Moreover, based on the results of Fig. \ref{xl3}, we can conclude the following remark that decreases the search space and processing time during the optimal design of the considered system. 

{\bf Remark 5.} {\it If $\sigma_{uqy}<\sigma_{uqx}$, then the optimal value for $N_{uqy}$ will be greater than the optimal value for $N_{uqx}$.}

\begin{figure}
	\centering
	\subfloat[] {\includegraphics[width=3 in]{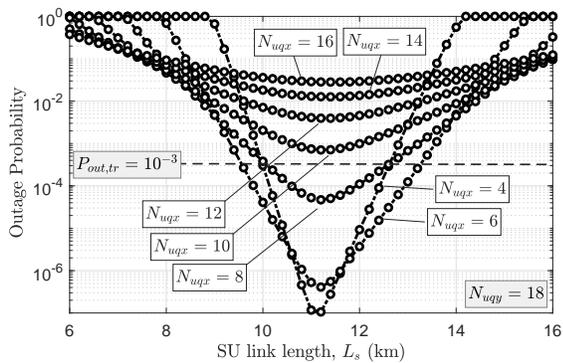}
		\label{ml1}
	}
	\hfill
	\subfloat[] {\includegraphics[width=3 in]{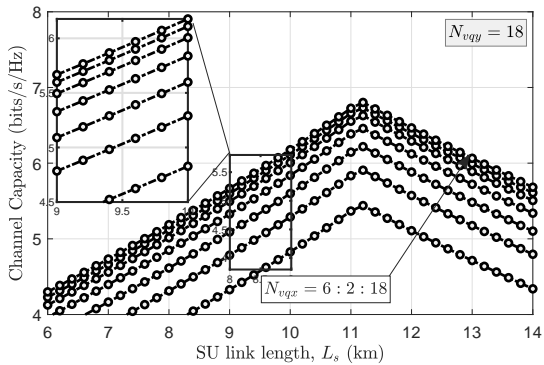}
		\label{ml2}
	}
	\caption{E2E performance of the considered system versus $L_s$ for different values of $N_{uqx}$ in terms of (a) outage probability and (b) channel capacity.}
	\label{ml3}
\end{figure}

In order to obtain more information about the optimal selection of $N_{uqx}$, in Figs. \ref{ml1} and \ref{ml2}, the outage probability and the channel capacity of the single relay system is plotted for different values of $N_{uqx}$. 
From the results of Fig. \ref{ml3}, although the channel capacity increases with increasing $N_{uqx}$, the antenna beam bandwidth decreases for large $N_{uqx}$ and the system becomes more sensitive to alignment errors. Therefore, as we observe, the channel capacity is maximized 
for $N_{uqx}=16$ and 18. However, for those values of $N_{uqx}$, we have $\mathbb{P}_\text{out}>\mathbb{P}_\text{out,tr}$ for all values of $L_s$ 
and therefore, the required QoS in condition \eqref{e3} is not guaranteed.
It seems that the optimal value for $N_{uqx}$ is equal to 14. 
However, it should be noted that the optimal value for $N_{uqx}$ cannot be determined from the results of Figs. \ref{xl3} and \ref{ml3}. According to constraint \eqref{e3}, in the entire flight path of the UAV, which is characterized by the parameter $-\pi<\theta_{R1}<\pi$, we should have $\mathbb{P}_\text{out}>\mathbb{P}_\text{out,tr}$.
%


\begin{figure}
	\centering
	\subfloat[] {\includegraphics[width=3 in]{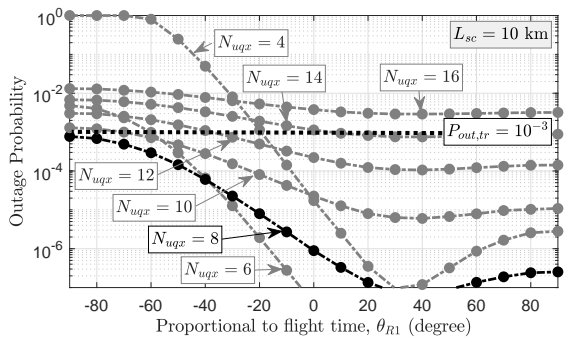}
		\label{xm1}
	}
	\hfill
	\subfloat[] {\includegraphics[width=3 in]{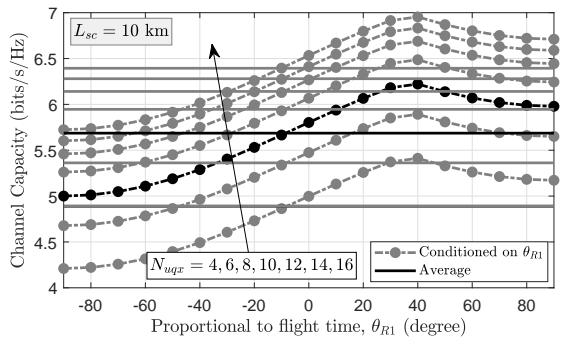}
		\label{xm2}
	}
	\caption{E2E performance of the considered system versus $\theta_{R1}$ for $L_{sc}=10$ km and different values of $N_{uqx}$ in terms of (a) outage probability and (b) channel capacity.}
	\label{xm3}
\end{figure}

\begin{figure}
	\centering
	\subfloat[] {\includegraphics[width=3 in]{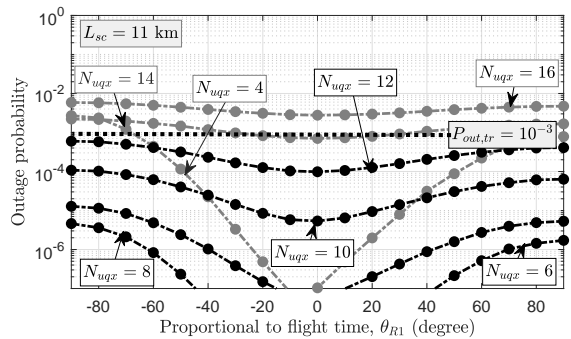}
		\label{xn1}
	}
	\hfill
	\subfloat[] {\includegraphics[width=3 in]{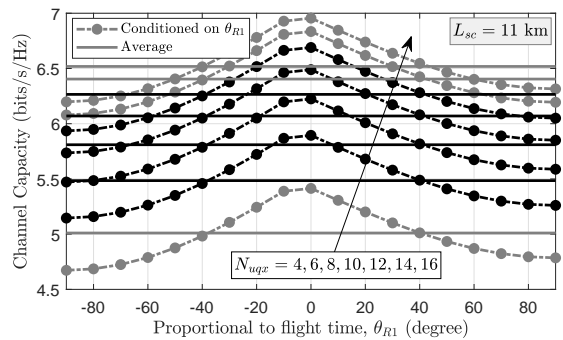}
		\label{xn2}
	}
	\caption{E2E performance of the considered system versus $\theta_{R1}$ for $L_{sc}=11$ km and different values of $N_{uqx}$ in terms of (a) outage probability and (b) channel capacity.}
	\label{xn3}
\end{figure}

\begin{figure}
	\centering
	\subfloat[] {\includegraphics[width=3 in]{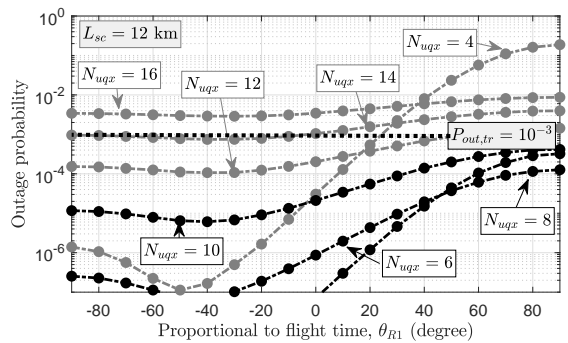}
		\label{xf1}
	}
	\hfill
	\subfloat[] {\includegraphics[width=3 in]{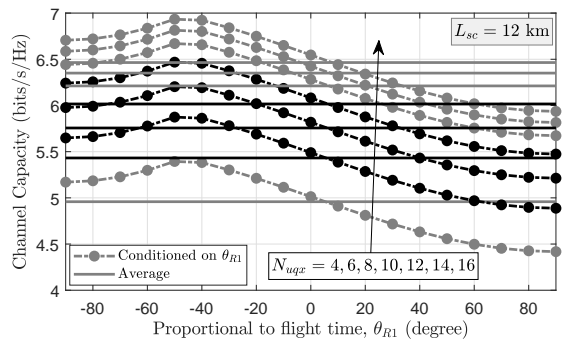}
		\label{xf2}
	}
	\caption{E2E performance of the considered system versus $\theta_{R1}$ for $L_{sc}=12$ km and different values of $N_{uqx}$ in terms of (a) outage probability and (b) channel capacity.}
	\label{xf3}
\end{figure}

Accordingly, in Figs. \ref{xm3}-\ref{xf3}, the end-to-end performance of the single relay system is examined along the entire flight path. Since the circular flight path is symmetric with respect to $\theta_{R1}$, the outage probability and the channel capacity are provided for interval $-\frac{\pi}{2}<\theta_{R1}<\frac{\pi}{2}$ instead of interval $-\pi<\theta_{R1}<\pi$.
Another important point is that the position of the circular flight path is controlled by the adjustable parameter $L_{sc}$ and has a very important impact on the system performance. To get a better understanding, the results of Figs. \ref{xm3}, \ref{xn3}, and \ref{xf3} are obtained for the three values of $L_{sc}=10$, 11, and 12 km, respectively.
Based on \eqref{opt1}, we seek to maximize the channel capacity while ensuring the constraints of \eqref{opt1}, especially, constraint \eqref{e3}. In this work we consider $\mathbb{P}_\text{out,tr}=10^{-3}$.
In these figures, in addition to the channel capacity, the average channel capacity over the entire circular flight path is also presented. Based on the results of Figs. \ref{xm3}-\ref{xf3}, by increasing $N_{uqx}$, the average channel capacity increases. However, by increasing $N_{uqx}$, the performance of the considered system in terms of outage probability is not necessarily improved. Only those $N_{uqx}$ values that achieve outage probability lower than $10^{-3}$ for the whole interval $-\frac{\pi}{2}<\theta_{R1}<\frac{\pi}{2}$ are acceptable and are marked in black color in those figures. For the other $N_{uqx}$ values, which in part or for the whole circular route can not guarantee the constraint of \eqref{e3}, we have marked in gray color. For instance, for $L_{sc}=10$ km, $N_{uqx}=8$ only guarantees \eqref{e3}, and therefore the maximum average channel capacity available for it is 5.3 bit/s/Hz. 
By changing the UAV's circular route from $L_{sc}=10$ to 12 km, we see that for $N_{uqx}=6$, 8, 10, and 12, condition \eqref{e3} is guaranteed. Then, among the values of $N_{uqx}=6$, 8, 10, and 12, it is observed that $N_{uqx}=12$ has the highest average capacity $\bar{\mathbb{C}}_{e2e}=6.6$ bit/s/Hz. 
As a result, a very important point that can be deduced from the results of Figs. \ref{xm3}-\ref{xf3} is provided in the following remark.

{\bf Remark 6.} {\it To calculate the optimal UAV's flight path as well as the optimal value for $L_{sc}$, it is better to first investigate the performance of the considered system versus $L_s$ since it gives a better view in terms of finding the acceptable interval for $L_s$ and $L_d$.
In other words, by doing this we will find an acceptable range of $L_{sc}$ and thus, the search space to find the optimal values of system parameters will be significantly reduced. Then for each value of $L_{sc}$, we find the values of $N_{uqx}$ that guarantee constraint \eqref{e3}, and the largest of the obtained $N_{uqx}$ results in the maximum achievable average channel capacity. This process will be repeated for all possible values of $L_{sc}$.}

\subsection{Multi-Relay Case}
The use of a single relay system can ultimately guarantee a certain length of $L_{sd}$. For example, for the parameters presented in Table \ref{I2}, the maximum possible length for $L_{sd}$ is 18.3 km, and for links longer than $L_{sd}>18.3$ km, two UAVs should be used. The results obtained so far for a single relay system are basic information for designing a two-relay or multi-relay system, but they are not sufficient. In addition to the parameters considered for a single relay system, for a two-relay system, it is necessary to find the optimal distance between the UAVs as well as the optimal antenna pattern used for communication links between the UAVs, which is studied in the following.
The parameters used to simulate the two-relay system are similar to the parameters considered for the single-relay system, except that for the two-relay system the length $L_{sd}$ has been increased to 25 km.

\begin{figure}
	\centering
	\subfloat[] {\includegraphics[width=3 in]{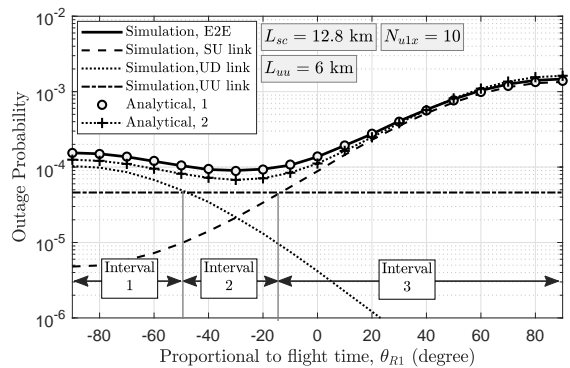}
		\label{nc1}
	}
	\hfill
	\subfloat[] {\includegraphics[width=3 in]{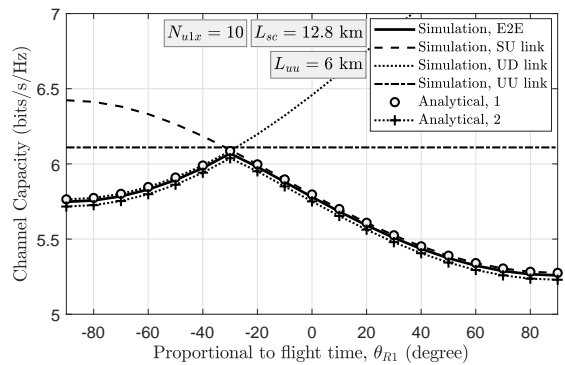}
		\label{nc2}
	}
	\caption{E2E performance of two-relay system versus $\theta_{R1}$ and comparison with the performance of CU, UU, and UD links in terms of (a) outage probability and (b) channel capacity.}
	\label{nc3}
\end{figure}

In Fig. \ref{nc3}, we evaluate the performance of a two-relay system in terms of both outage probability and channel capacity versus $\theta_{R1}$. Based on Eqs. \eqref{g1} and \eqref{g2}, the performance of the system depends on the performance of the three SU, UU, and UD links. To get a better understanding, the performance of each link is also provided separately. As it turned out, for lower values of $L_s$ obtained for the interval close to $\theta_{R1}=-90^o$, the system performance is limited to UD link. In Fig. \ref{nc1}, we have specified this interval with the name of interval 1. Then, for the intermediate values of $\theta_{R1}$, the system performance is limited to the inter-UAV link or UU link marked with interval 2. For larger values of $\theta_{R1}$, the length of $L_s$ increases and thus, the system performance is limited to SU link determined by interval 3 in Fig. \ref{nc1}. In addition, the accuracy of the analytical expressions has been confirmed using Monte-Carlo simulations. In Fig. \ref{nc3}, the label Analytical 1 refers to the analytical expression obtained based on the channel distribution function derived in \eqref{sm3}, and the label Analytical 2 refers to the analytical expression obtained based on the channel distribution function provided in \eqref{sm8}.
As can be seen, Analytical 1 is more accurate. However, the results of Analytical 2 have less computational load than Analytical 1.
%

\begin{figure}
	\centering
	\subfloat[] {\includegraphics[width=3 in]{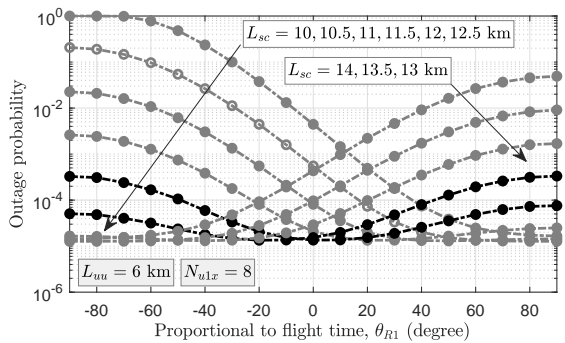}
		\label{mc1}
	}
	\hfill
	\subfloat[] {\includegraphics[width=3 in]{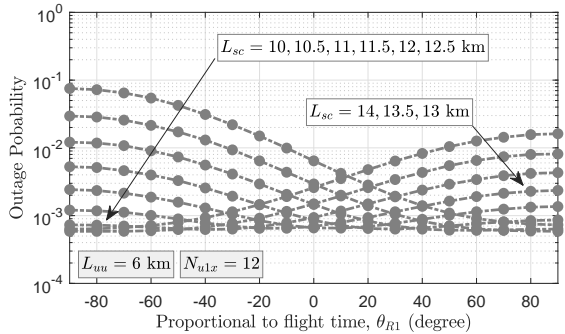}
		\label{mc2}
	}
	\caption{Outage probability of two-relay system versus $\theta_{R1}$ for $L_{uu}=6$ km, different values of $L_{sc}$, and (a) $N_{tx1}=N_{rx2}=8$ and (b) $N_{tx1}=N_{rx2}=12$.}
	\label{mc3}
\end{figure}

\begin{figure}
	\centering
	\subfloat[] {\includegraphics[width=3 in]{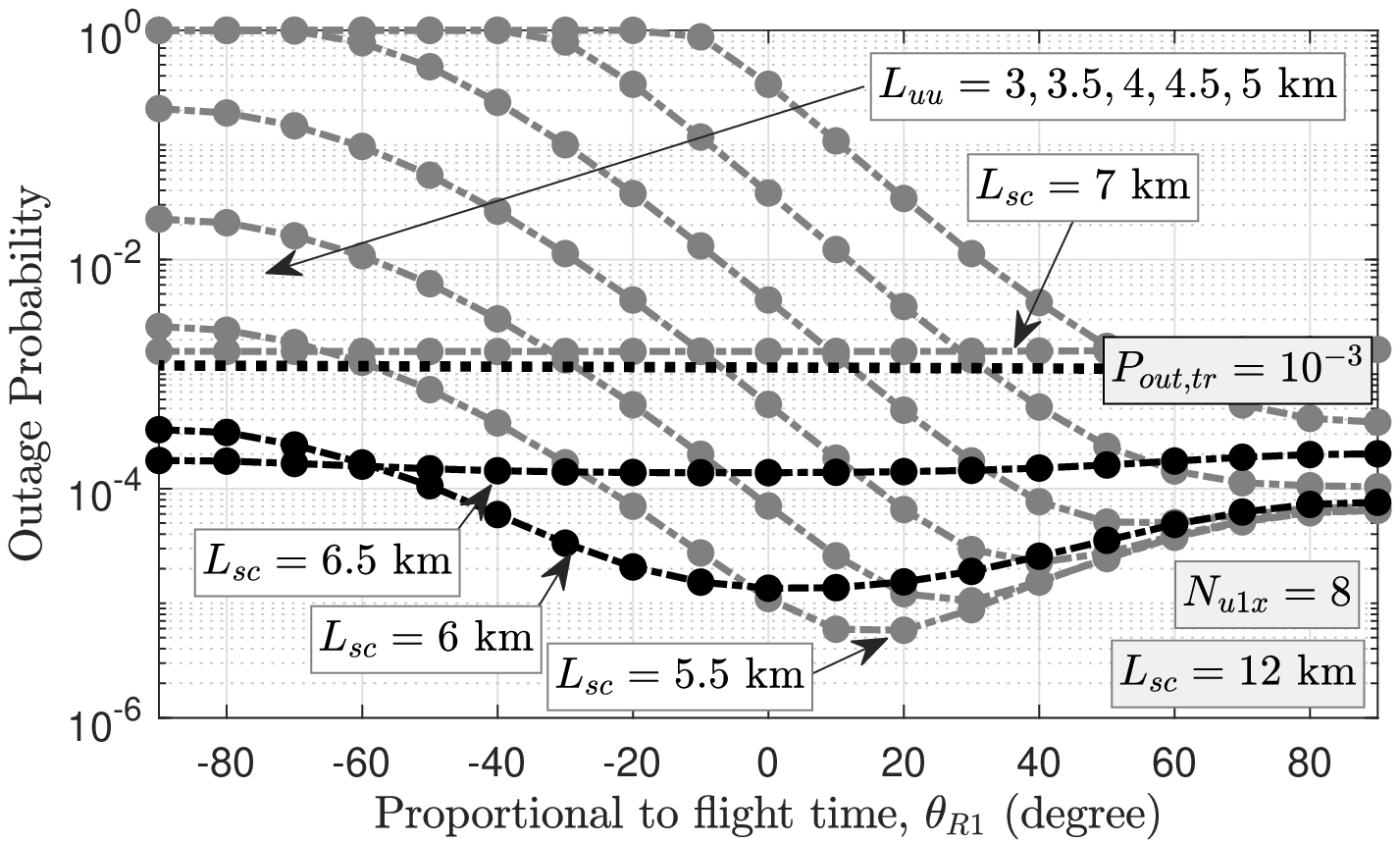}
		\label{fc1}
	}
	\hfill
	\subfloat[] {\includegraphics[width=3 in]{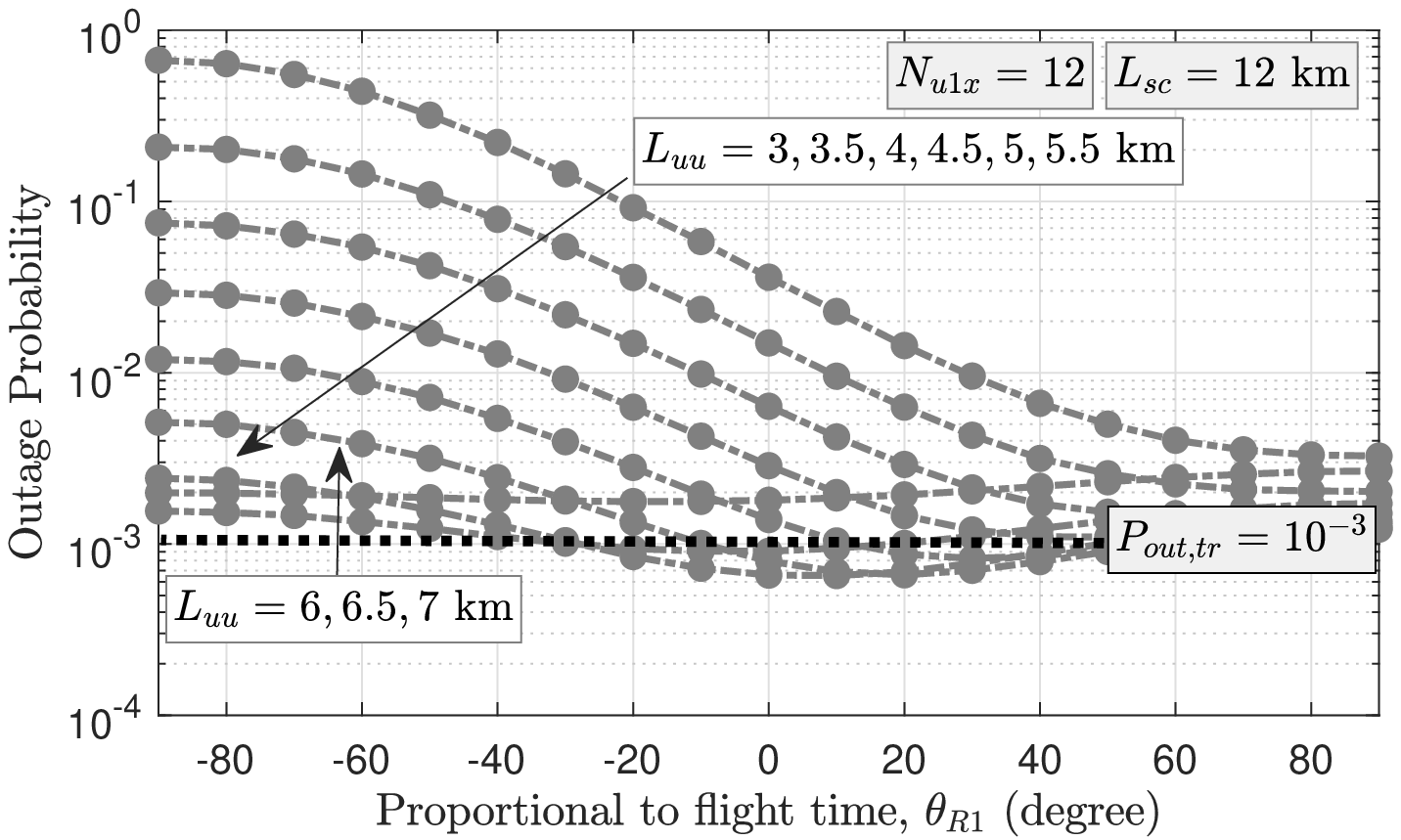}
		\label{fc2}
	}
	\caption{Outage probability of two-relay system versus $\theta_{R1}$ for $L_{sc}=12$ km, different values of $L_{uu}$, and (a) $N_{tx1}=N_{rx2}=N_{u1x}=8$ and (b) $N_{tx1}=N_{rx2}=12$.}
	\label{fc3}
\end{figure}

Note that the results of Fig. \ref{nc3} are obtained for given the values of $N_{tx1}=N_{rx2}=10$, $L_{uu}=6$ km, and $L_{sc}=12$ km. However, those parameters are adjustable and by changing them, the performance of the considered system changes, significantly. 
Accordingly, in Fig. \ref{mc3}, outage probability of a two relay system for different values of $L_{sc}$ is plotted. The results of this figure clearly show the importance of finding an optimal value for $L_{sc}$.
This indicates that for smaller values of $L_{sc}$, the system performance is limited to the UD link, and as $L_{sc}$ increases, the link length of $L_d$ becomes shorter, resulting in improved outage probability of the UD link.
For intermediate values of $L_{sc}$, the outage probability on the circular flight path has less changes, indicating that the system performance is limited by the UU link because the UU link has a fixed link length along the circular flight path.
For longer values of $L_{sc}$, it is observed that by increasing $\theta_{R1}$, the outage probability significantly increases. In this case, the system performance is limited to the SU link. In other words, it indicates that in half of the UAV's circular path, the system performance has an acceptable outage probability, and in the other half, $L_s$ increases and the system performance is significantly reduced. 
Moreover, the results of Fig. \ref{mc1} are provided for $N_{tx1}=N_{rx2}=8$ and the results of Fig. \ref{mc2} are for $N_{tx1}=N_{rx2}=12$. It is observed that by changing $N_{tx1}$, the system performance as well as the optimal values for $L_{sc}$ also change.
In addition, the results of Fig. \ref{mc3} are obtained for a constant value $L_{uu}$. To show the importance of finding an optimal value for $L_{uu}$, outage probability of the considered system is provided in Fig. \ref{fc3} for different value of $L_{uu}$. As the results of those figures show, similar to the $L_{sc}$, by changing the values of $L_{uu}$, outage probability of the considered two-relay system changes, significantly.

\subsection{Optimal System Design}
Finally, in order to optimally design the tunable parameters of the considered two-relay system, the method adopted in Tables \ref{Ir1} and \ref{Ir2} can be used. For example, in Table \ref{Ir1}, for a given value of $L_{sc}=12$ km, we calculate the outage probability of the considered system for different values of $N_{tx1}$ and $L_{uu}$. Note that the outage probability does not need to be calculated on the entire flight path. It is enough to calculate on the critical points $\theta_{R1}=-\frac{\pi}{2}$ and $\theta_{R1}=\frac{\pi}{2}$. If constraint \eqref{e3} is met at those points, constraint \eqref{e3} will be met for the entire flight path and is shown in Tables \ref{Ir1} and \ref{Ir2} with a ``+" sign. Then, for the points that guarantee constraint \eqref{e3}, we calculate the average channel capacity and select the values for $N_{tx1}$ and $L_{uu}$ that result in the highest average channel capacity.
For example, according to the results in Table \ref{Ir1}, although the maximum average channel capacity is $\bar{\mathbb{C}}_{e2e}=6.54$ bit/s/Hz for $L_{uu}=6.5$ km and $N_{tx1}=12$, it does not guarantee constraint \eqref{e3} along the entire flight path.
Therefore, for $L_{sc}=12$ km, the maximum achievable channel capacity is $\bar{\mathbb{C}}_{e2e}=5.72$ bit/s/Hz that will be obtained for $L_{uu}=6.5$ km and $N_{tx1}=10$. 
Note that the values of $L_{uu}=6.5$ km and $N_{tx1}=10$ are optimal only for $L_{sc}=12$ km and the results should be repeated for the rest of the probable values of $L_{sc}$. Finally, we select the optimal value from the entire search space.

\section{Conclusions}
By taking into account the actual channel parameters such as the UAV vibrations, tracking error, real 3GPP antenna pattern, UAV's height and flight path, and considering the effect of physical obstacles, the optimal design of a relay system based on fixed wing UAV was investigated. In particular, we derived the distribution of SNR which is based on the sum of a series of  Dirac delta functions.
Then, we used the SNR distribution and derived the closed-form expressions for the outage probability and the channel capacity of the considered system as a function of all real system parameters. 
After that, we extended the analytical expressions for a multi-relay system. 
The accuracy of closed-form expressions was verified with the results obtained from Monte-Carlo simulations.
Finally, by providing sufficient simulation results, we investigate the effects of key channel parameters such as antenna pattern gain and optimal flight path on the performance of the considered system and we carefully study the relationships between these parameters in order to maximize average channel capacity.

It is necessary to note a few points. Even if the tunable parameters of the considered system are optimally designed for a specific geographical area, the optimal values must be constantly updated. For example, the performance of the considered UAV-assisted system is highly dependent on weather conditions, especially wind speed. Because the wind speed is changing during the day and night, the UAVs' instabilities change and as a result, it is expected that the optimal value for the antenna pattern and UAVs' position will change.
Since the use of Monte-Carlo simulation is time consuming, given the high accuracy, we hope that the provided analytical expressions will help to analyze and design of the considered system with high accuracy, more easily and in a shorter time.

\begin{table}
	\caption{Comparison of the optimal values for $N_{tx1}=N_{rx2}=N_{u1x}$, and $L_{uu}$ to achieve maximum average channel capacity to guarantee $\mathbb{P}_\text{out}<10^{-3}$ over the whole circular flight path when $L_{sc}=12$ km.} 
	\centering 
	\begin{tabular}{c } 
		\includegraphics[width=2.5 in]{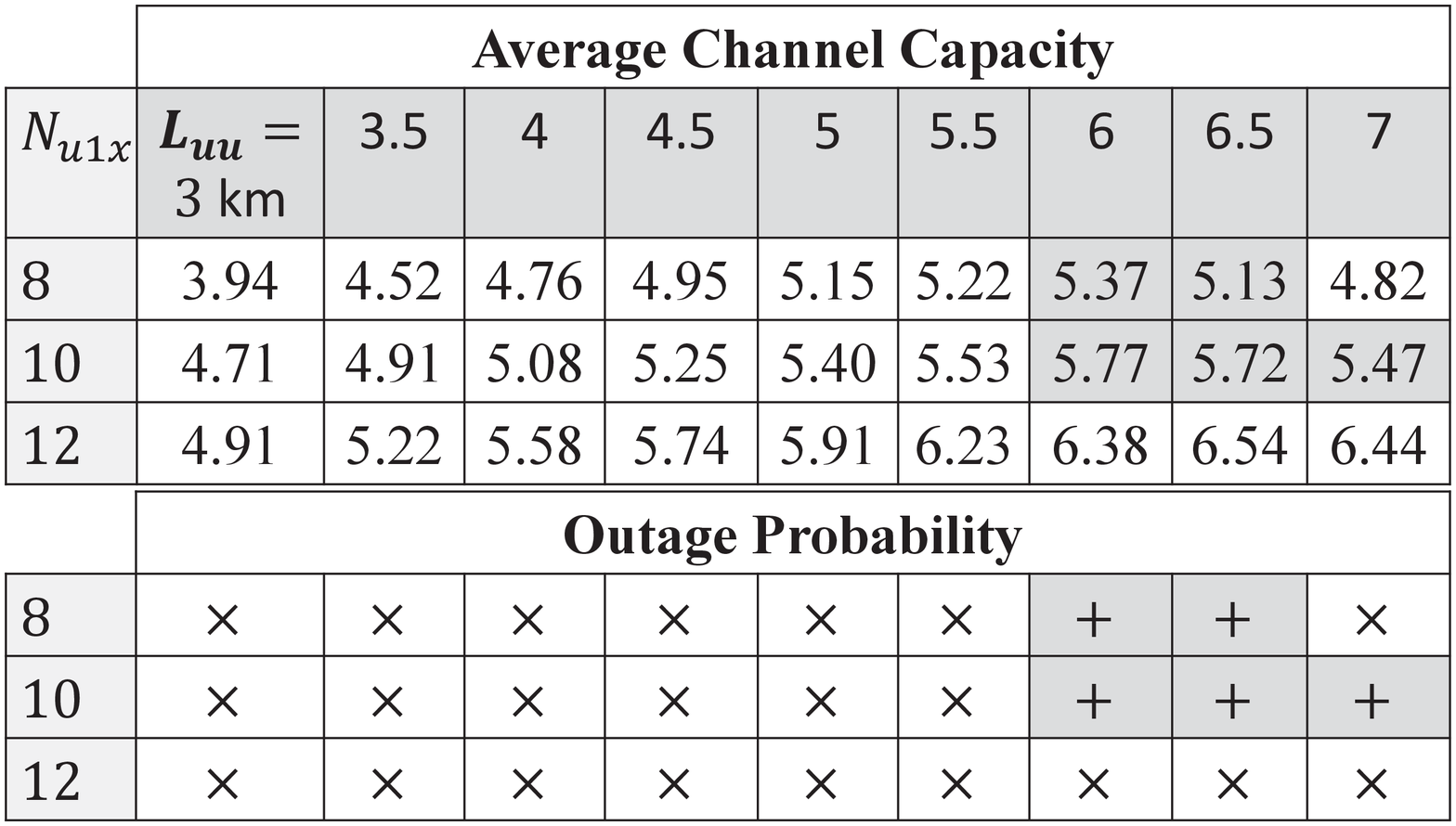}     
	\end{tabular}
	\label{Ir1} 
\end{table}

\begin{table}
	\caption{Comparison of the optimal values for $N_{tx1}=N_{rx2}=N_{u1x}$, and $L_{sc}$ to achieve maximum average channel capacity to guarantee $\mathbb{P}_\text{out}<10^{-3}$ over the whole circular flight path when $L_{uu}=6$ km.} 
	\centering 
	\begin{tabular}{c } 
		\includegraphics[width=2.5 in]{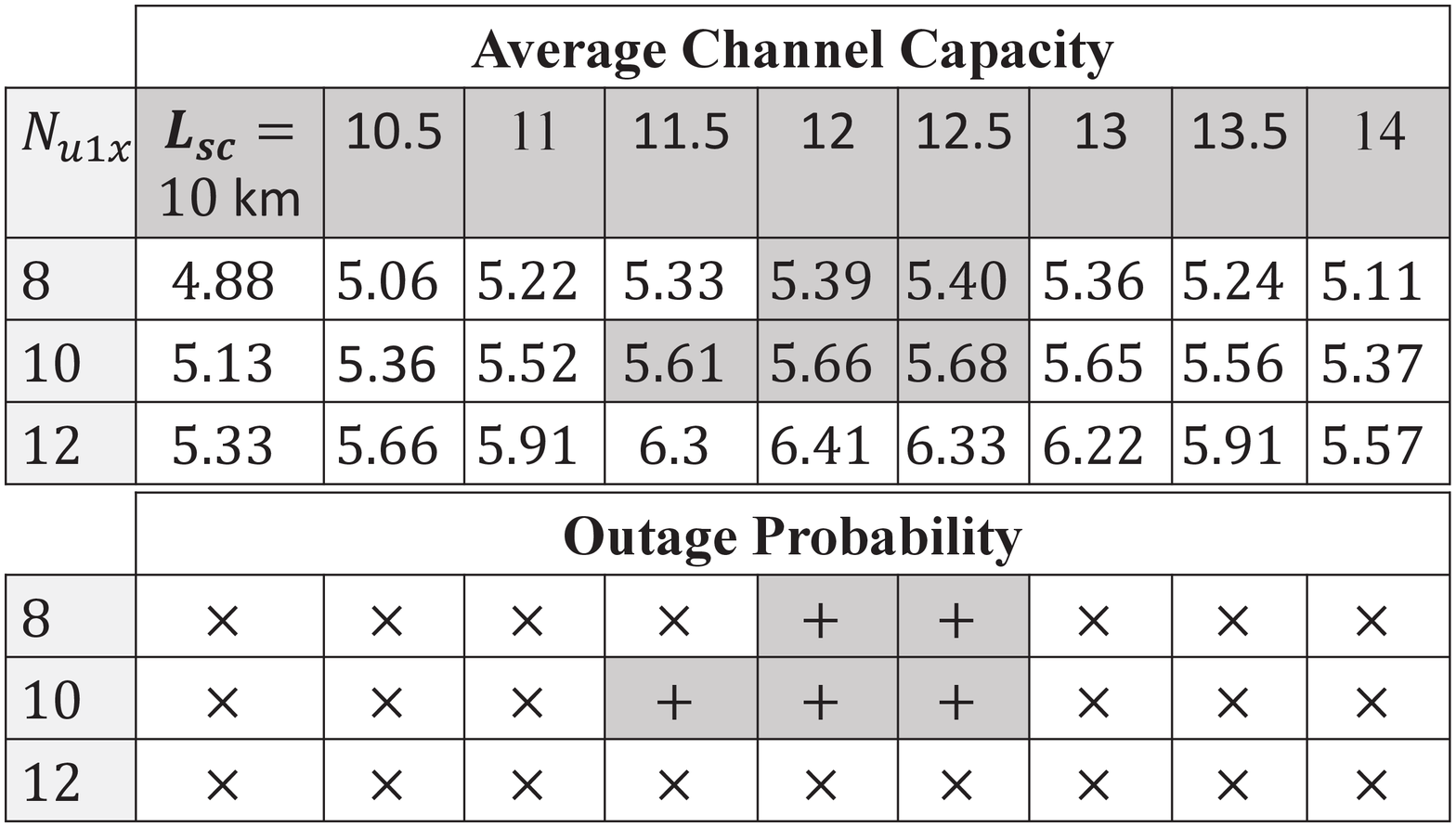}     
	\end{tabular}
	\label{Ir2} 
\end{table}

%
%
%

\appendices

\section{SNR Distribution}
\label{AppA}
Due to the lower changes of the considered antenna's gain pattern in the Roll direction, for the lower antenna misalignment (less than a few degrees), \eqref{ro} can be approximated with good accuracy as follows:
\begin{align}
	\label{sm1}
	&\Gamma_{q}(\theta_{qxy},\theta_{uqxy}|\theta_{R1} ) \simeq \Gamma_{q1}(\theta_{R1},N_{qw},N_{uqw}) \nonumber  \\
	& \times
	\left( \frac{\sin\left(\frac{N_{qx} k d_{qx} 
			\sin\left(\theta_{qxy}\right)
		}{2}\right)} 
	{N_{qx}\sin\left(\frac{k d_{qx} 
			\sin\left( \theta_{qxy} \right)
		}{2}\right)}
	\frac{\sin\left(\frac{N_{uqx} k d_{uqx} 
			\sin\left(\theta_{uqxy}\right)
		}{2}\right)} 
	{N_{uqx}\sin\left(\frac{k d_{uqx} \sin\left(\theta_{uqxy}\right)
		}{2}\right)}
	\right)^2,
\end{align}
where
\begin{align}
	\label{pu1}
	&\Gamma_{q1}(\theta_{R1},N_{qw},N_{uqw}) = \frac{P_{t,q} h_L(L_q(\theta_{R1},L_{sc},H_u)) }{\sigma_n^2}  
	\nonumber \\
	&\times 10^{G_\text{max}/5} G_0(N_{qx},N_{qy}) G_0(N_{uqx},N_{uqy}),
\end{align}
and $\theta_{qxy}=\sqrt{\theta_{qx}^2 + \theta_{qy}^2}$, and $\theta_{uqxy}=\sqrt{\theta_{uqx}^2 + \theta_{uqy}^2}$. Since we have $\theta_{qx}\sim\mathcal{N}(\mu_{qx},\sigma_{qx}^2)$ and 
$\theta_{qy}\sim\mathcal{N}(\mu_{qy},\sigma_{qy}^2)$, the random variable $\theta_{qxy}$ follows  the Beckmann distribution as \cite{pena2017generalized}
\begin{align}
	\label{k1}
	&f_{\theta_{qxy}}(\theta_{qxy}) = 
	\frac{\theta_{qxy}}{2\pi \sigma_{qx} \sigma_{qy}}  \int_0^{2\pi} 
	\exp\left(- \frac{(\theta_{qxy}\cos(\theta)-\mu_{qx})^2} {2 \sigma_{qx}^2} \right. \nonumber \\
	& \left.~~~~~~~~~~~~~~~~
	- \frac{(\theta_{qxy}\cos(\theta)-\mu_{qy})^2} {2 \sigma_{qy}^2}
	\right) \text{d}\theta.
\end{align} 
Similarly, the distribution of RV $\theta_{uqxy}$ is obtained from \eqref{k1} by substituting $\mu_{uqx}$, $\mu_{uqy}$, $\sigma_{uqx}$, and $\sigma_{uqy}$ instead of  $\mu_{qx}$, $\mu_{qy}$, $\sigma_{qx}$, and $\sigma_{qy}$, respectively.
Let us approximate \eqref{sm1} as 
\begin{align}
	\label{sm2}
	&\Gamma_{q}(\theta_{qxy},\theta_{uqxy}|\theta_{R1} ) \simeq \frac{\Gamma_{q1}(\theta_{R1},N_{qw},N_{uqw}) }
	{N_{qx}^2 N_{uqx}^2}
	\sum_{j_q =1}^{KJ_q} \sum_{j_u =1}^{KJ_u} \nonumber  \\
	&       
	\left( \frac{\sin\left(\frac{N_{qx} k d_{qx} 
			\sin\left(\frac{2j_q}{J_q N_{qx}}\right)
		}{2}\right)} 
	{\sin\left(\frac{k d_{qx} 
			\sin\left( \frac{2j_q}{J_q N_{qx}} \right)
		}{2}\right)} 
	  \frac{\sin\left(\frac{N_{uqx} k d_{uqx} 
			\sin\left(\frac{2j_u}{J_u N_{uqx}}\right)
		}{2}\right)} 
	{\sin\left(\frac{k d_{uqx} \sin\left(\frac{2j_u}{J_u N_{uqx}}\right)
		}{2}\right)}
	\right)^2 \nonumber \\
	& \times \left[ \mathbb{Y}\left(\theta_{uqxy}- \frac{2 (j_u-1)}{J N_{uqx}} \right)  -  
	\mathbb{Y}\left(\theta_{uqxy} - \frac{2 j_u}{J_u N_{uqx}} \right)  \right] \nonumber \\
	& \times \left[ \mathbb{Y}\left(\theta_{qxy}- \frac{2 (j_q-1)}{J_q N_{qx}} \right)  -  
	\mathbb{Y}\left(\theta_{qxy} - \frac{2 j_q}{J_q N_{qx}} \right)  \right],
\end{align}
where 
$
\mathbb{Y}(x)= \left\{
\begin{array}{rl}
	1& ~~~ {\rm for}~~~ x\geq 0 \\
	0& ~~~ {\rm for}~~~ x< 0 \\
\end{array} \right. 
$ is the sign function,
and the parameters $J_q$, $J_u$, and $K$ are the natural numbers that for large values of $J_q$ and $J_u$, \eqref{sm2} tends to \eqref{sm1}.
Also, $K=1$ refers to the main lobe of the antenna pattern and $K>1$ refers to the number of sidelobes. 
Using Eqs. \eqref{k1}, \eqref{sm2}, and \cite[(17)-(20)]{zhu2017distribution}, and after some derivations, the distribution of $\Gamma_{q}(\theta_{qxy},\theta_{uqxy} )$ conditioned on $\theta_{R1}$ is derived in \eqref{sm3}.

	\section{SNR Distribution provided in Proposition 1}
\label{AppB}
The Beckmann distribution given in \eqref{k1} can be approximated as \cite{boluda2016novel}
\begin{align}
	\label{k2}
	&f_{\theta_{qxy}}(\theta_{qxy}) \simeq 
	\frac{\theta_{qxy}}{\sigma_{qc}^2 }  
	\exp\left(- \frac{\theta_{qxy}^2} {2 \sigma_{qc}^2} 
	\right),
\end{align} 
where
\begin{align} 
	\label{k3}
	\sigma_{qc}^2 = \left( \frac{3\mu_{qx}^2\sigma_{qx}^4 + 3\mu_{qy}^2\sigma_{qy}^4 + \sigma_{qx}^6 + \sigma_{qy}^6 }
	{2} \right)^{\frac{1}{3}}.
\end{align}
Similarly, the distribution of RV $\theta_{uqxy}$ is obtained from \eqref{k2} by substituting $\mu_{uqx}$, $\mu_{uqy}$, $\sigma_{uqx}$, and $\sigma_{uqy}$ instead of  $\mu_{qx}$, $\mu_{qy}$, $\sigma_{qx}$, and $\sigma_{qy}$, respectively.
Using Eqs. \eqref{sm2}, \eqref{k3}, and \cite[eq. (3.321.4)]{jeffrey2007table}, and after some derivations, the distribution of $\Gamma_{q}(\theta_{qxy},\theta_{uqxy} )$ conditioned on $\theta_{R1}$ is derived in \eqref{sm7}.

\section{Channel Capacity}
\label{AppC}
We approximate the average end-to-end channel capacity conditioned on $\theta_{R1}$ provided in \eqref{c7} as
\begin{align}
	\label{df5}
	\mathbb{C}_{e2e|\theta_{R1}} = \min\left\{
	\mathbb{C}_{su}(\theta_{R1}),
	\mathbb{C}_{du}(\theta_{R1})		\right\},
\end{align}
where $\mathbb{C}_{su}(\theta_{R1})$ and $\mathbb{C}_{du}(\theta_{R1})$ are the average channel capacities of SU and UD links conditioned on $\theta_{R1}$, respectively. For our system model, $\mathbb{C}_{qu}(\theta_{R1})$ is a function of random variables (RVs) 
$\theta_{qx}$, $\theta_{qy}$, $\theta_{uqx}$, and $\theta_{uqy}$ and can be obtained as 
\begin{align}
		\label{c4}
		&\mathbb{C}_{qu}(\theta_{R1}) = 
		\frac{1}{4\pi^2 \sigma_{qx} \sigma_{qy} \sigma_{rqx} \sigma_{rqy} }
		\int_0^{\pi/2} \int_0^{\pi/2} \int_0^{\pi/2} \int_0^{\pi/2} \nonumber \\
		&\log_2\left(1 + \Gamma_{q}(\theta_{qx},\theta_{qy},\theta_{uqx},\theta_{uqy}|\theta_{R1} ) \right) 
		\exp\left( -\frac{(\theta_{qx}-\mu_{qx})^2}{2\sigma_{qx}}  \right) \nonumber \\
		& \times
		\exp\left( -\frac{(\theta_{qy}-\mu_{qy})^2}{2\sigma_{qy}}  \right)
		\exp\left( -\frac{(\theta_{uqx}-\mu_{uqx})^2}{2\sigma_{uqx}}  \right)
		\nonumber \\
		&\times \exp\left( -\frac{(\theta_{uqy}-\mu_{uqy})^2}{2\sigma_{uqy}}  \right)
		\text{d}\theta_{qx}  \text{d}\theta_{qy}
		\text{d}\theta_{uqx}  \text{d}\theta_{uqy}.
\end{align}
where $\Gamma_{q}(\theta_{qx},\theta_{qy},\theta_{uqx},\theta_{uqy}|\theta_{R1} )$ is obtained in \eqref{ro}.
Although the expression given in \eqref{c4} reduces the 8-dimensional integral to 4-dimensional, it still has a high computational time. Using Eqs. \eqref{sm3}, \eqref{df5}, and \eqref{c4} and performing a series of calculations, the end-to-end channel capacity is derived in \eqref{sn7}.
Also, using Eqs. \eqref{sm7}, \eqref{df5}, and \eqref{c4} and performing a series of calculations, another closed-form expression for $\mathbb{C}_{qu|\theta_{R1}}$ with lower computational load is derived in  \eqref{sn9}.

\section{Outage Probability}
\label{AppD}
We consider that the NFP use DF relay system. 
Outage probability of considered system conditioned on $\theta_{R1}$ is obtained as:
\begin{align}
	\label{e6}
	\mathbb{P}_{\text{out}|\theta_{R1}} &= \text{Prob}\Big\{ 
	\min\Big[ \Gamma_{s}(\theta_{sx},\theta_{sy},\theta_{usx},\theta_{usy}|\theta_{R1}), \nonumber \\
	&~~~\Gamma_{d}(\theta_{dx},\theta_{dy},\theta_{udx},\theta_{udy}|\theta_{R1} ) \Big]
	<\Gamma_\text{th}\Big\} \nonumber \\
	%
	&=1-\text{Prob}\Big\{ 
	\min\Big[ \Gamma_{s}(\theta_{sx},\theta_{sy},\theta_{usx},\theta_{usy}|\theta_{R1}), \nonumber \\
	&~~~\Gamma_{d}(\theta_{dx},\theta_{dy},\theta_{udx},\theta_{udy}|\theta_{R1} ) \Big]
	>\Gamma_\text{th}\Big\} \nonumber \\
	&=1-\text{Prob}\Big\{ 
	\Big[ \Gamma_{s}(\theta_{sx},\theta_{sy},\theta_{usx},\theta_{usy}|\theta_{R1}),~\& \nonumber \\
	&~~~\Gamma_{d}(\theta_{dx},\theta_{dy},\theta_{udx},\theta_{udy}|\theta_{R1} ) \Big]
	>\Gamma_\text{th}\Big\}, 
\end{align}
where $\Gamma_\text{th}$ is the SNR threshold. Since the random variables $\theta_{sx}$, $\theta_{sy}$, $\theta_{usx}$, and $\theta_{usy}$ are independent of the random variables 
$\theta_{dx}$, $\theta_{dy}$, $\theta_{udx}$, and $\theta_{udy}$, therefore \eqref{e6} can be rewritten as follows:
\begin{align}
	\label{e7}
	\mathbb{P}_{\text{out}|\theta_{R1}} &= 1-\text{Prob}\Big\{ 
	\Gamma_{s}(\theta_{sx},\theta_{sy},\theta_{usx},\theta_{usy}|\theta_{R1}) >\Gamma_\text{th}\Big\}\nonumber \\
	&~~~\times \text{Prob}\Big\{ 
	\Gamma_{d}(\theta_{dx},\theta_{dy},\theta_{udx},\theta_{udy}|\theta_{R1}) >\Gamma_\text{th}\Big\} \nonumber \\
	&= \mathbb{P}_{\text{out,su}|\theta_{R1}} + \mathbb{P}_{\text{out,du}|\theta_{R1}} - \mathbb{P}_{\text{out,su}|\theta_{R1}}\mathbb{P}_{\text{out,du}|\theta_{R1}}
\end{align}
where $\mathbb{P}_{\text{out,su}|\theta_{R1}}$ and $\mathbb{P}_{\text{out,du}|\theta_{R1}}$ are the outage probability of SU and UD links, respectively. 
Using Eqs. \eqref{sm2} and \eqref{e7}, the end-to-end outage probability of single relay system is derived in Proposition 3.




\end{document}